%% file: main.tex
\documentclass[a4paper,11pt]{article}
\usepackage{jheppub} 
\usepackage{lineno}
\usepackage{placeins}
\input{definitions}

\usepackage{slashed}
\usepackage{booktabs}

\title{\boldmath Charming Darwin: the evolution of QCD parameters across different species}

\author[a]{F. Bernlochner}
\author[b]{A. Gilman}
\author[b]{S. Malde}
\author[a]{M. Prim}
\author[c,d]{K.K. Vos}
\author[b]{G. Wilkinson}

\affiliation[a]{Physikalisches Institut, University of Bonn, Nußallee 12 53115 Bonn, Germany}
\affiliation[b]{Department of Physics, University of Oxford, Keble Road, Oxford OX1 3RH, United Kingdom}
\affiliation[c]{Gravitational Waves and Fundamental Physics (GWFP), Maastricht University, Duboisdomein 30, NL-6229 GT Maastricht, the Netherlands}
\affiliation[d]{Nikhef, Science Park 105, NL-1098 XG Amsterdam, the Netherlands}
\emailAdd{alexander.leon.gilman@cern.ch}

\abstract{
We explore the prospects of measurements of spectral moments of inclusive charm decays with BESIII. The rich and uniquely clean data set of charm mesons and baryons at BESIII offers a unique laboratory to study the evolution of Heavy Quark Expansion (HQE) parameters across different charm hadron species and to shed light on the interplay between heavy quark dynamics and light quark effects. The HQE in terms of inverse powers of the heavy meson mass is well-established in beauty decays, however due to the lighter charm mass its applicability to charm remains an open question. To date no determination of the HQE parameters, the kinetic energy, chromomagnetic moment and Darwin terms, has been attempted. A particularly important role here is given to the Darwin and weak annihilation operators, whose values are important to predict lifetimes of heavy hadrons. Using a fast simulation for the BESIII detector response and predictions for spectral moments, we investigate the sensitivity to HQE parameters with today's and possible future data sets. In addition, we discuss the theory challenges for the HQE in charm and the experimental limitations. We further investigate the sensitivity of determining the CKM matrix element $|V_{cs}|$ with inclusive semileptonic charm decays. 
}

\begin{document}
\maketitle
\flushbottom

\section{Introduction}

The Heavy Quark Expansion (HQE) is a powerful technique to study the dynamics of inclusive heavy hadron decays and predict their lifetimes. In this formalism, the decay rates of heavy quarks can be related to the forward scattering amplitude and the time ordered products can be expanded using the Operator Product Expansion (OPE). The first non-trivial contribution to the decay rate arises from dimension-5 operators (at order $1/m_Q^2$), where $m_Q$ is the mass of the heavy quark and the non-perturbative matrix elements $\{\mu_\pi^2, \mu_G^2\}$ correspond to the kinetic energy and the chromomagnetic moment of the heavy quark inside the heavy meson. Dimension-6 operators (entering at order $1/m_Q^3$) result in two additional contributions $\{ \rho_D^3, \rho_{\mathrm{LS}}^3 \}$ corresponding to the Darwin and spin-orbit terms. The Darwin operator provides one of the key corrections to the free heavy quark decay lifetime~\cite{Lenz:2020oce} and the HQE is constructed in such a way that it depends on the flavour of the initial heavy hadron. 

For inclusive semileptonic beauty meson decays, the HQE has achieved remarkable precision, with hadronic matrix elements identified up to dimension 8 \cite{Mannel:2023yqf} and $\alpha_s^3$ corrections to the total decay rate calculated \cite{Fael:2020tow}, collected in the open-source package Kolya \cite{Fael:2024fkt}. This progress, combined with experimental precision, enables the extraction of inclusive $|V_{cb}|$ at the percent level (see, e.g., \cite{Bernlochner:2022ucr,Finauri:2023kte} and references therein). Furthermore, the possibility of measuring inclusive $B_s$ decays using a sum-over-exclusive technique has been studied at the LHC~\cite{DeCian:2023ezb}.

For charm decays, the challenge is the low mass of the charm which lies close to the non-perturbative QCD regime, which may lead to a break down of the OPE. For example, four-quark operators, weak annihilation, and Pauli interference terms enter in the HQE for charm meson lifetimes. These terms are numerically enhanced by $16\pi^2$ and play an important role in explaining the large lifetime differences in charm decays \cite{Lenz:2013aua,Lenz:2014jha,Belle-II:2021cxx}. 

On the other hand, the semileptonic widths of the different charm decays are more similar, which may indicate that weak annihilation (WA) operators are of less importance. This gives confidence in the validity of the OPE, which was discussed in ~\cite{Voloshin:2002je,Falk:1995kn}, with a recent reevaluation in \cite{Fael:2019umf}. 
Precisely due to the slower convergence, inclusive charm decays are significantly more sensitive to HQE hadronic matrix elements, such as $\rho_D^3$, as well as to WA operators. These WA operators are crucial input parameters for Standard Model predictions of $B\to X_u \ell \bar\nu_\ell$ ~\cite{Ligeti:2010vd,Becirevic:2008us,Gambino:2010jz,Bigi:2009ym} and $B\to X_{s,d}\ell \ell$ \cite{Huber:2015sra,Huber:2019iqf,Huber:2020vup} decays. As discussed above they also form crucial inputs in lifetime predictions~\cite{Albrecht:2024oyn}. Charm meson and baryon lifetimes have recently been revisited and measured with thus far unprecedented precision by Belle II~\cite{Belle-II:2021cxx,Belle-II:2022ggx,Belle-II:2023eii,Belle-II:2022plj}. 

In this paper, we explore the opportunities that measurements of inclusive semileptonic charm decays could provide. We assess the potential of the large data set recorded by the Beijing Electron Spectrometer Mk. III (BESIII), which is particularly intriguing due to its clean samples of $D^0$, $D^+$, $D_s^+$, and $\Lambda_c^+$ decays. Such an experimental program, combined with corresponding theoretical progress, would permit a detailed study of the evolution of non-perturbative QCD parameters across various species of heavy hadrons, including the first determination of these parameters in heavy baryon decays. In addition, this program would, for the first time, link semileptonic charm hadron measurements with precision tests of the Standard Model through charm hadron lifetimes.

Previous measurements of inclusive semileptonic $D^0$, $D^+$, and $D_s^+$ decays at CLEO \cite{CLEO:2009uah} have already been used to determine the magnitude of weak annihilation operators \cite{Ligeti:2010vd,Gambino:2010jz}. More recently, measurements from BESIII \cite{BESIIIDsIncSL, BESIII:2022cmg} have been used to determine the strength of the strong coupling constant \cite{Wu:2024jyf} at the charm scale.

In Ref. \cite{Gambino:2010jz}, CLEO's differential measurements of lepton energies were converted into moments. When combined with the HQE parameters obtained from $B \to X_c \ell \bar{\nu}_\ell$ inclusive decays, this enabled the extraction of weak annihilation effects, which were found to be small, suggesting that the HQE may converge sufficiently rapidly. We propose a dedicated analysis of the lepton energy moments and the as-yet unexplored di-lepton invariant mass moments of $D^0$, $D^+$, $D_s^+$, and $\Lambda_c^+$ decays. This approach could pave the way for a full determination of charm HQE parameters directly from charm data. The comparison between $D$ and $D_s^+$ decays would allow for an $SU(3)$ symmetry test based on data, providing crucial inputs for lifetime determinations \cite{Lenz:2020oce}. Moreover, comparing these findings with the HQE parameters obtained from inclusive $B$ decays will be intriguing. Finally, combining these results with branching ratio measurements \cite{CLEO:2009uah,BESIIIDsIncSL,BESIII:2022cmg} could facilitate the extraction of both $|V_{cs}|$ and $|V_{cd}|$ in a manner analogous to the inclusive $|V_{cb}|$ extraction \cite{Bernlochner:2022ucr,Bordone:2021oof,Finauri:2023kte}, with the potential to achieve uncertainties competitive with exclusive determinations.

The remainder of this manuscript is organized as follows: Section~\ref{sec:Sensitivity} introduces a realistic fast simulation of BESIII for semileptonic decays, including estimates for systematic uncertainties. Section~\ref{sec:calib} discusses the experimental approach to calibrating detector-level observables to unfolded raw spectral moments. Section~\ref{sec:theory} reviews the HQE for charm and details the challenges associated with the lighter charm quark compared to beauty decays. Section~\ref{sec:extraction} presents an exploratory study on extracting the experimental precision of HQE parameters from the unfolded spectral moments. Lastly, Section~\ref{sec:outlook} provides a detailed outlook on the prospects for the inclusive charm program outlined in this work.

\section{Prospects for inclusive semileptonic measurements at BESIII}
\label{sec:Sensitivity}

BESIII records electron-positron collisions provided by the Beijing Electron-Positron Collider Mk. II (BEPCII) \cite{BEPCII}. BEPCII can deliver center-of-mass energies between 1.85 and 4.95 GeV and the BESIII experiment's large data sets at open-charm-hadron pair-production thresholds provide ideal data samples to perform inclusive semileptonic measurements in the charm system. From the interior to the exterior, the BESIII detector~\cite{BESIIIDetector} is comprised of a drift chamber tracking system, a plastic scintillator time-of-flight system, a crystal calorimeter, a 1T super-conductor solenoid, and a resistive-plate-chamber muon system. The drift chamber provides sub-percent level momentum resolution of charged particles and, combined with measurements from the time-of-flight system, is able to distinguish well between protons, charged pions, and charged kaons. The BESIII calorimeter allows for excellent identification of electrons and provides percent-level resolution on energy depositions of charged and neutral particles. 

The sensitivities to HQE parameters from measurements of \DpIncSL, \DzIncSL, \DsIncSL, and \LcIncSL at BESIII are estimated by analyzing Monte Carlo simulation produced with the EvtGen package~\cite{EvtGen}. A measurement strategy similar to that employed in inclusive semileptonic beauty decays at Belle and Belle II \cite{BelleQ2,BelleIIQ2} is assumed, in which the lepton and all hadrons accompanying the lepton in the final state of the semileptonic decay are reconstructed. This differs from techniques applied in previous measurements of inclusive semileptonic charm decays \cite{CLEO:2009uah,BESIIIDsIncSL, BESIII:2022cmg}, in which the hadronic system $X$ was not reconstructed and only the charged lepton was identified. These measurements produced precise determinations of the inclusive branching fractions and the electron momentum spectra in the laboratory frame. With additional assumptions on the production mechanism of charm-hadron pairs, the measured laboratory-frame distributions allowed for HQE analysis of the \Dz, \Dp, and \Dsp decays to constrain the contribution of weak-annihilation operators  \cite{Gambino:2010jz,Ligeti:2010vd}. 
In this study, we examine how precisely the explicit reconstruction of the hadronic $X$ system can be performed to construct Lorentz-invariant observables. Such quantities are crucial for mitigating the proliferation of HQE parameters at higher orders~\cite{Fael:2018vsp,Mannel:2018mqv}. This reconstruction enables the reconstruction of the Lorentz-invariant lepton-neutrino mass squared, $q^2$, and the Lorentz-invariant mass of the hadronic system, $M_X$. Additionally, properties of the $X$ system allow for the separation of $c \to s$ and $c \to d$ transitions, facilitating the first extraction of the CKM matrix elements $|V_{cs}|$ and $|V_{cd}|$ from inclusive semileptonic charm decays.

In the explored analysis procedure, double-tag techniques similar to those developed by the MARKIII collaboration~\cite{MARKIII_one,MARKIII_Two} are employed with the tag decay modes\footnote{Charge conjugation is implied here and throughout the paper.} $\Dzb\to K^-\pi^+$, $\Dm\to K^-\pi^+\pi^+$, from Ref.~\cite{BESIII:2023cym} for $\Dsm$,  and from Ref.~\cite{BESIII:2022cmg} for $\Lcm$. The dataset listed for BESIII analysis of \Dsp mesons primarily contains \Dsp mesons produced through $e^+e^-\to D_s^{*\pm}D_s^{\mp}$, and so the identification of a $D_s^{*}$ meson through a $D_s^{*}\to D_s \gamma$ is also assumed, where the $D_s$ meson produced through the $D_s^{*}$ can decay to either a semileptonic decay or to a hadronic tag final-state. The assumed size of BESIII data samples are listed Table~\ref{tab:BESIIILumi}. The listed data sets correspond to samples which BESIII has analyzed for publication, with the exception of the samples for \Dz and \Dp analysis, where a larger dataset has recently been collected.

\begin{table}[htbp]
    \caption{Integrated luminosities, center-of-mass energies and estimated inclusive semileptonic double-tag yields of BESIII data sets assumed in sensitivity studies.}
    \centering
    \begin{tabular}{c| c|c|c|c}
         &\Dz  & \Dp  & \Dsp \cite{Ds_Lumi} & \Lcp \cite{Lc_Lumi1,Lc_Lumi2}  \\ \hline
         \Ecm $[\GeV]$   & 3.773 & 3.773 & 4.130--4.230  & 4.600-4.699\\\hline
     Integrated Luminosity $[\invfb]$ &  21  & 21 & 7.1  &4.5\\\hline
     Estimated Double-tag Yields &  200000   & 700000 & 30000 & 4300 \\\hline
          
    \end{tabular}
    \label{tab:BESIIILumi}
\end{table}

\subsection{Estimated distributions from fast simulation}
\label{sec:fastsim}

A fast-simulation software based on EvtGen~\cite{EvtGen} is developed to estimate the effects of BESIII detector response, momentum resolution, and geometric and kinematic acceptance on inclusive semileptonic measurements. The effects of final-state radiation are neglected in the simulation. The decays of charm hadrons are generated by EvtGen for the tag-side meson and all observed semileptonic decays \cite{PDG2022}, and those predicted by isotopic symmetry based on observed decays. The ISGW2 \cite{ISGW2} model is assumed for all $\{D^+, D^0, D_s^+, \Lambda_c^+ \} \to M e^+ \nu$ decays, where $M$ is any hadron, and a uniform phase-space model is assumed when more than one hadron is produced in the semileptonic decay. Detector effects are incorporated probabilistically based on publicly reported reconstruction efficiencies and resolutions. The efficiencies of tracking charged particles and particle identification are estimated based on the published measurement of the branching fraction of \DsIncSL decays \cite{BESIIIDsIncSL}. Photon reconstruction efficiencies and resolution are estimated based on Ref.~\cite{BESIIINeutrals}. The energy deposition of \KL mesons and neutrons in the BESIII calorimeter is estimated based on the analysis of \KL backgrounds in the search for $D^0\to\pi^0\nu\overline \nu$ at BESIII \cite{BESIIIPiNuNu}. The reconstruction efficiency and momentum resolution of $K_S^0\to \pi^+\pi^-$ decays is estimated based on Ref.~\cite{BESIIIIncLKS}. The efficiency and momentum resolution of reconstructing $\Lambda\to p\pi^-$ decays is assumed to be identical to $K_S^0\to \pi^+\pi^-$ decays due to the similar vertex reconstruction.

\begin{figure}[b!]
    \centering
    \begin{tabular}{cc}
  \includegraphics[width=7cm]{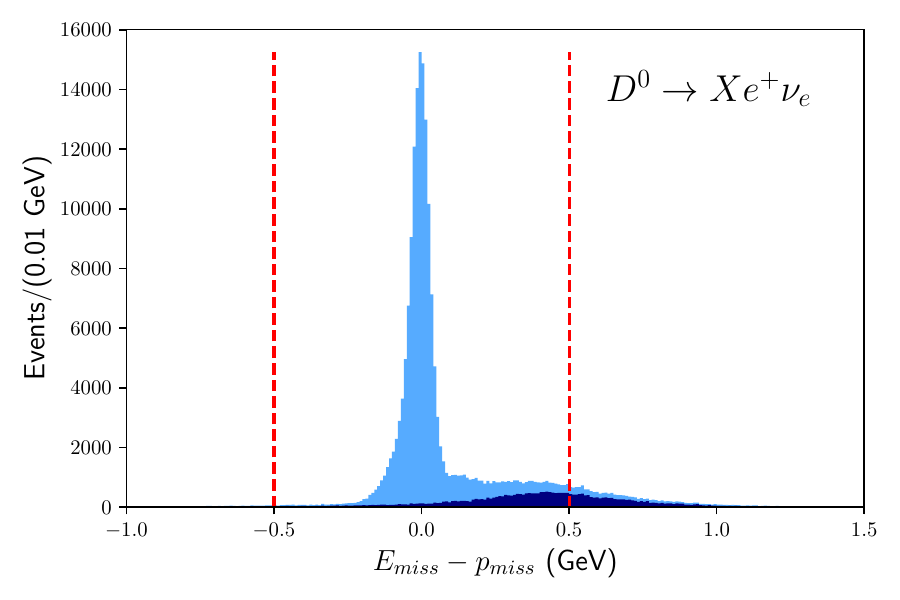}  & \includegraphics[width=7cm]{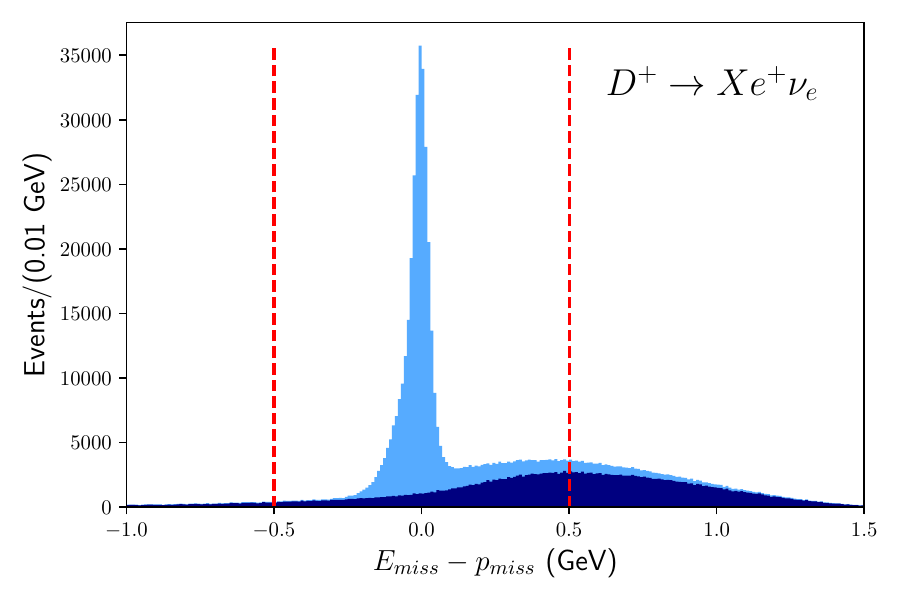} \\
  \includegraphics[width=7cm]{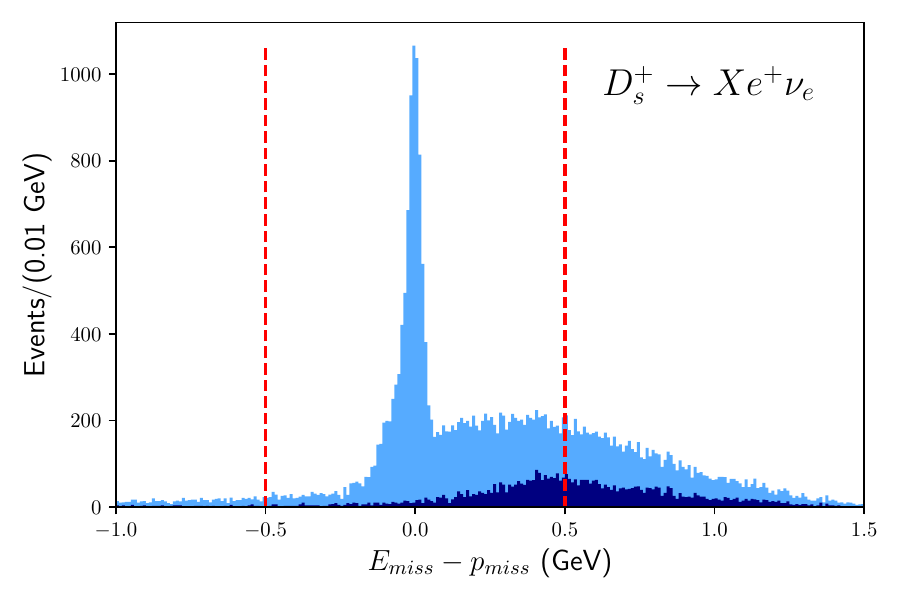} &
  \includegraphics[width=7cm]{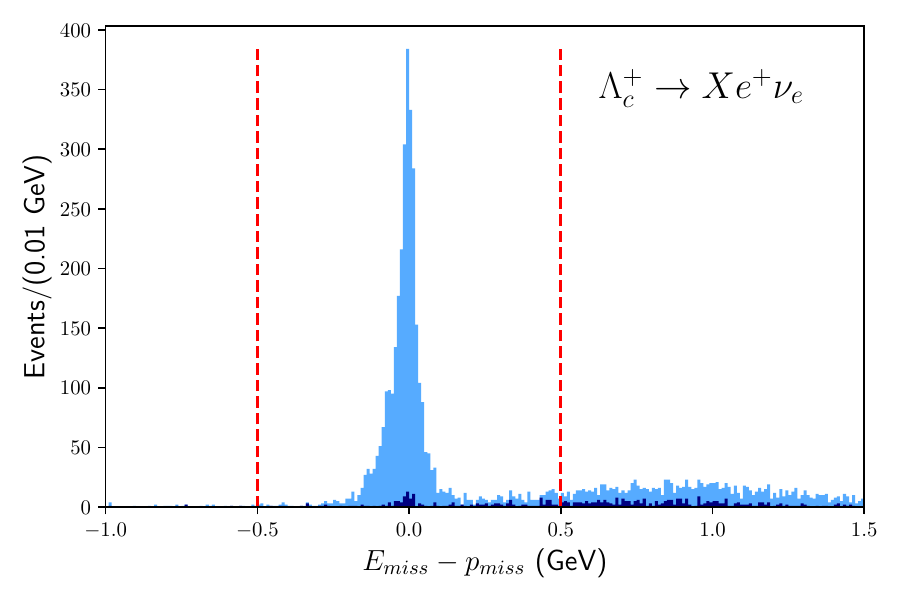} \\
  \end{tabular}
  
    \caption{ Distribution of the reconstructed \Empm for decays of (top-left) \DzIncSL, (top-right) \DpIncSL, (bottom-left) \DsIncSL, and (bottom-right) \LcIncSL produced by the fast simulation. The light blue histogram is the total distribution, and the dark blue histogram corresponds to events with \KL mesons, \KSToNeutrals decays, or neutrons in the final state. The \Empm selection requirements correspond to the interior of the dashed red lines.}
    \label{fig:DpEmPm}
\end{figure}

Simulated measured distributions of the kinematic distributions of interest in the HQE (the squared four-momentum of the $e^+\nu$ system \qSq, the $e^+$ energy in the parent rest frame \Elep, and the mass of the hadronic system in the final state \mX) are produced by the fast simulation based on the estimated double-tag sample sizes listed in Table~\ref{tab:BESIIILumi}. The BESIII detector is unable to consistently distinguish between electrons and other charged particles for candidate tracks with momenta less than 200\MeV. As such, it is required that the momentum of the electron candidate be above this threshold.  The resulting \Empm distribution produced by the fast simulation for each charm hadron is shown in Fig.~\ref{fig:DpEmPm}, with the contribution of decays with \KL mesons or \KSToNeutrals decays in the final state highlighted. As the figure shows, decays with \KL mesons or \KSToNeutrals decays are often poorly reconstructed, due to the poor resolution of \KL energy depositions in the calorimeter and the inability of the BESIII detector to distinguish between energy depositions due to photons and those due to \KL mesons. As such, a requirement of $\left|\Empm\right|<500\MeV$ is placed on simulated events to estimate the sensitivity of measurements at BESIII, and a correction is later applied to account for this requirement. It should be noted that a measurement at BESIII could correct for decays with \KL mesons or \KSToNeutrals decays in the final state that are removed with this selection  based on measurements of decays with \KSToPiPi decays in the final state, which BESIII reconstructs with high efficiency and good momentum resolution. Contributions from neutrons, primarily in $\Lambda_c^+$ decays, can similarly be corrected for assuming isotopic symmetry\footnote{The decay $\Lambda_c^+ \to n e^+ \nu$ must be handled separately.} based on observed decays with protons in the final state.

The decay $D_s^+\to \tau^+\nu_\tau, \tau^+\to e^+\nu_e\overline \nu_\tau$ is expected to pass the selection requirements at a similar rate to signal processes, as it did in the previous BESIII analysis of $D_s^+\to X e^+\nu$ \cite{BESIIIDsIncSL}. We suggest that it can be handled in a similar fashion in our procedure as the previous analysis, subtracting on its contribution based off of the measured branching fraction, as the modeling of this decay carries negligible systematic uncertainty.

\begin{figure}[hbtp]
    \centering
    \begin{tabular}{c}
    \begin{tabular}{cc}
    \includegraphics[width=7.25cm]{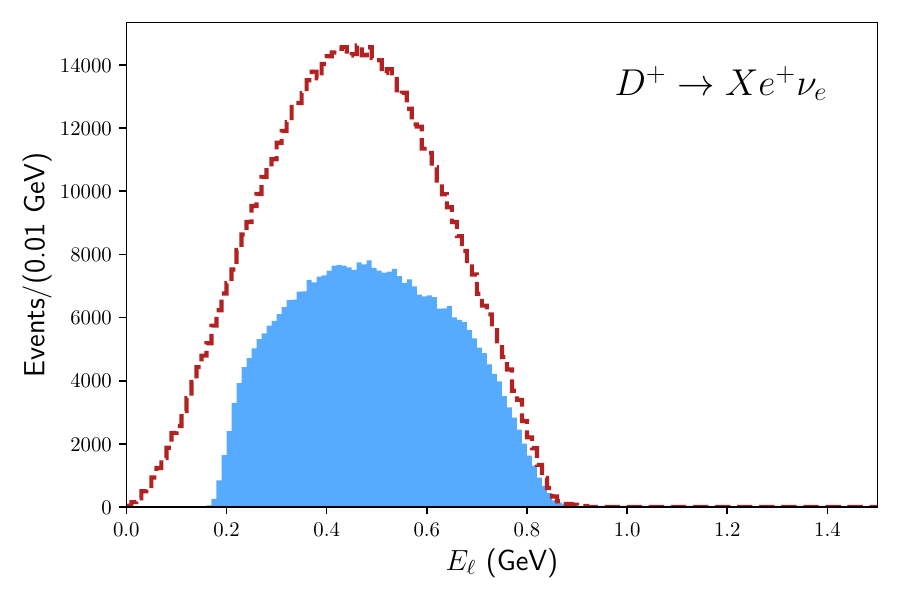} &\includegraphics[width=7.25cm]{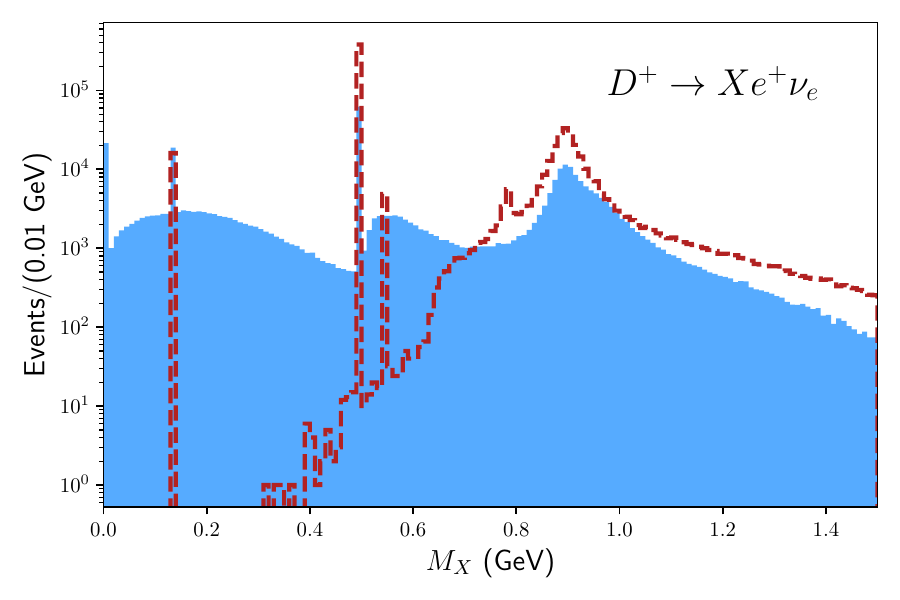} 
    \end{tabular}\\
    \includegraphics[width=7.25cm]{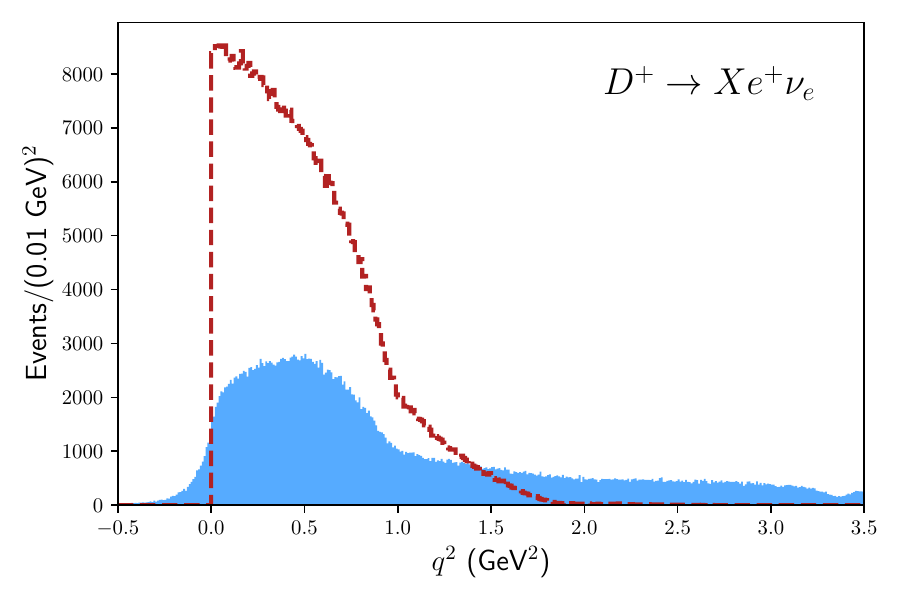} 
    \end{tabular}
    \caption{ Comparisons between the generated $\Dp$ distributions of the (top-left) \Elep, (top-right) \mX , and (bottom) \qSq variables and the reconstructed distributions produced by the fast simulation. In each plot, the generated distribution is shown as the dashed orange histogram and the reconstructed distribution is shown as the solid light blue histogram. }
    \label{fig:DpRecoDists}
\end{figure}

The kinematic and geometric acceptance of BESIII associated with reconstructing and identifying the positron and the \Empm requirement reduce the sample size for analysis, and sculpt the kinematic distributions of interest. The efficiency of the acceptance and selections is approximately 68\%, 53\%, 44\%, and 49\% for the \mbox{\DzIncSL}, \mbox{\DpIncSL}, \mbox{\DsIncSL}, and \mbox{\LcIncSL} samples, respectively. While the resolution is predicted to be considerably worse for the \Dsp sample, a measurement at BESIII could improve this with a kinematic constraint on the $D_s^{*}$ candidate. Comparisons between the generated kinematic distributions of interest of \DpIncSL decays and the corresponding  reconstructed distributions produced by the fast simulation are shown in Fig.~\ref{fig:DpRecoDists}.

\begin{figure}[t]
    \centering
\begin{tabular}{cc}
    \includegraphics[width=7.4cm]{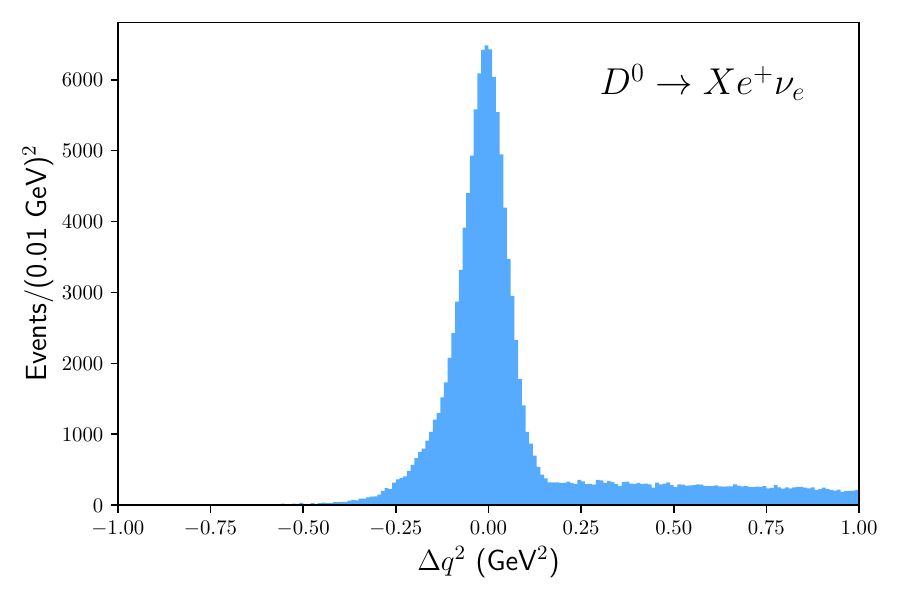} & \includegraphics[width=7.4cm]{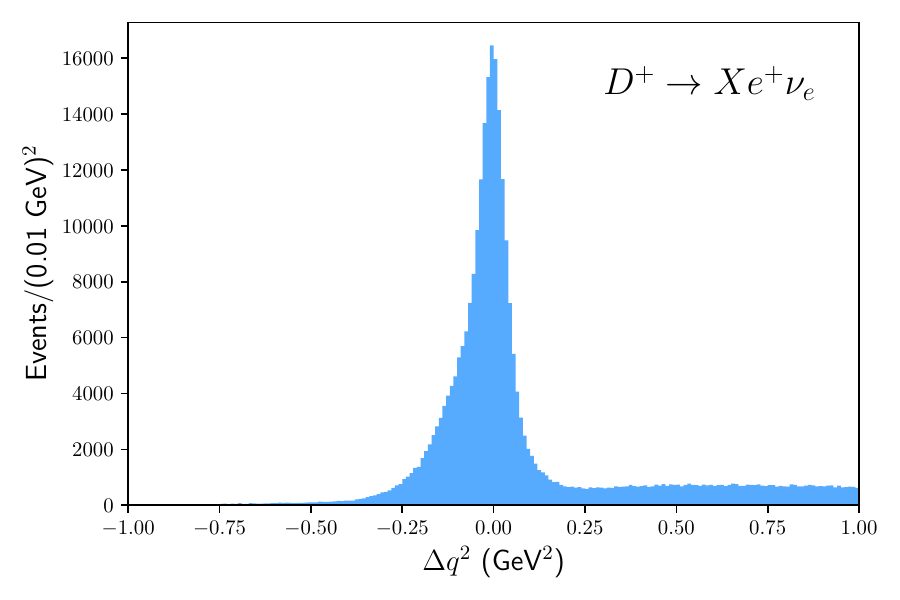}\\
    \includegraphics[width=7.4cm]{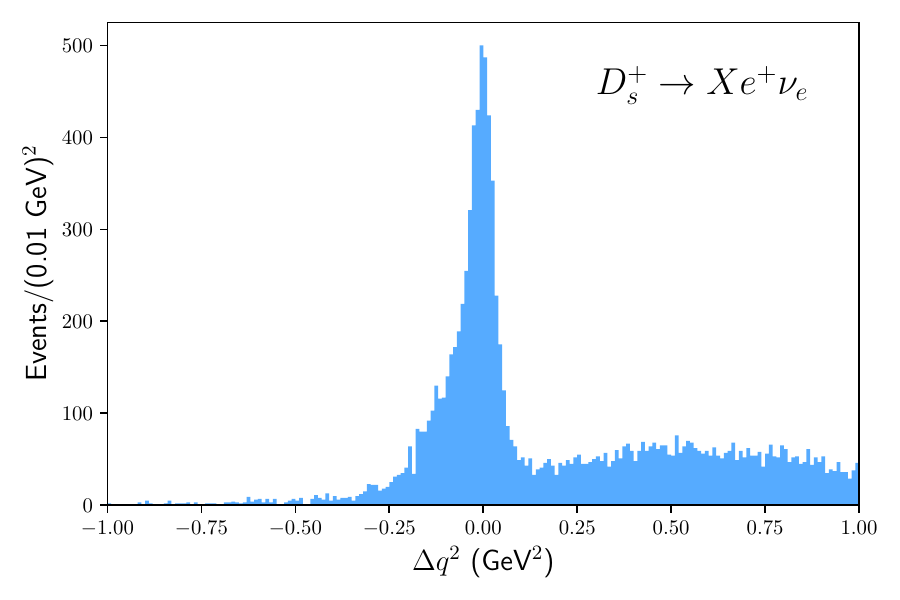} & \includegraphics[width=7.4cm]{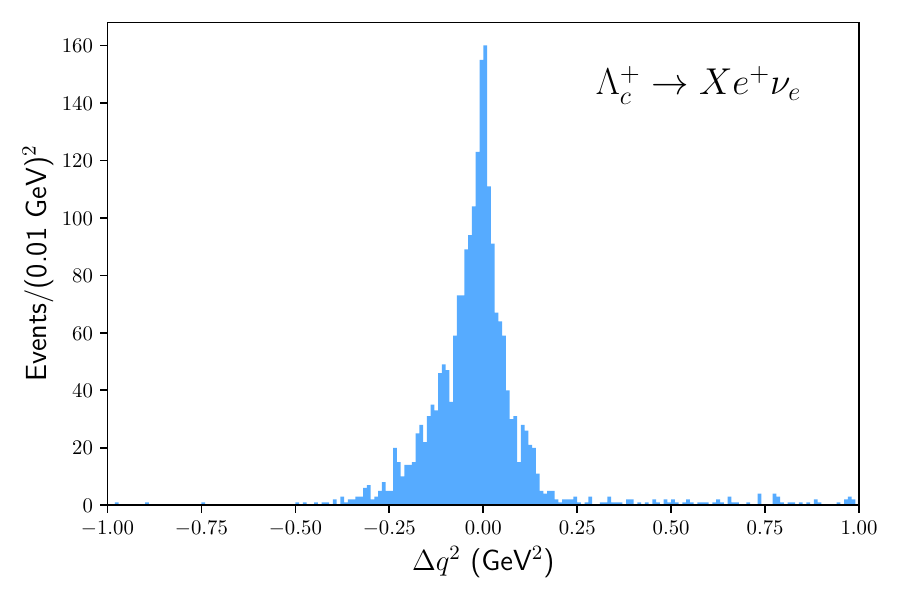}
\end{tabular}
    \caption{Predicted distribution of $\DqSq$ for events passing selections from (top-left) \DzIncSL, (top-right) \DpIncSL, (bottom-left) \DsIncSL, and (bottom-right) \LcIncSL.}
    \label{fig:q2dists}
\end{figure}

After the selection requirements are applied, the event-by-event difference between reconstruction and generated variables exhibit a clear core resolution with a tail due to missing particles in the reconstructed $X$ system. The effects of the BESIII resolution can be corrected for with a calibration procedure similar to Refs.~\cite{BelleQ2,BelleIIQ2}. The distribution of the difference of the reconstructed and generated $q^2$, referred to as \DqSq, is shown in Fig.~\ref{fig:q2dists}, where all discussed selections have been applied.

\begin{figure}
\centering
\begin{tabular}{cc}
    \includegraphics[width=0.4\linewidth]{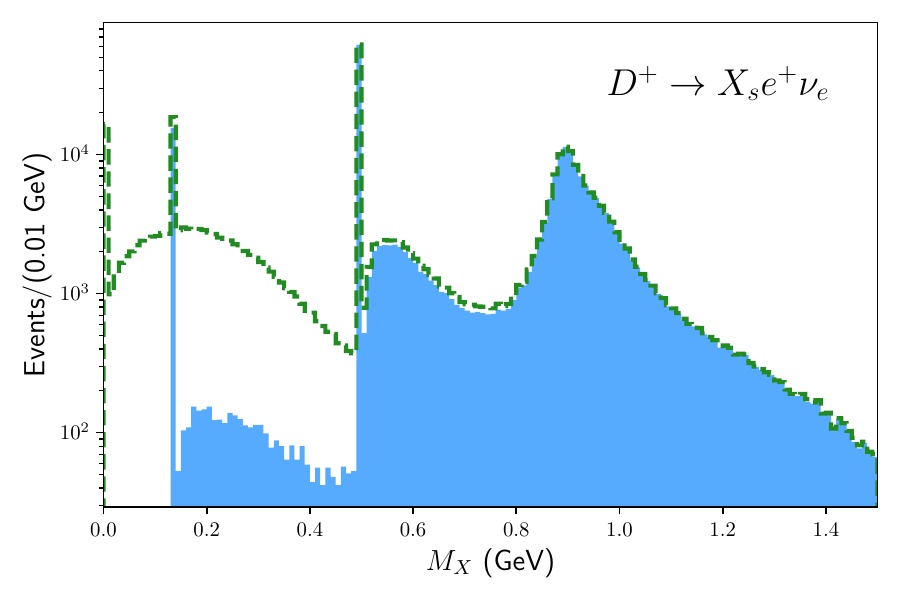} & \includegraphics[width=0.4\linewidth]{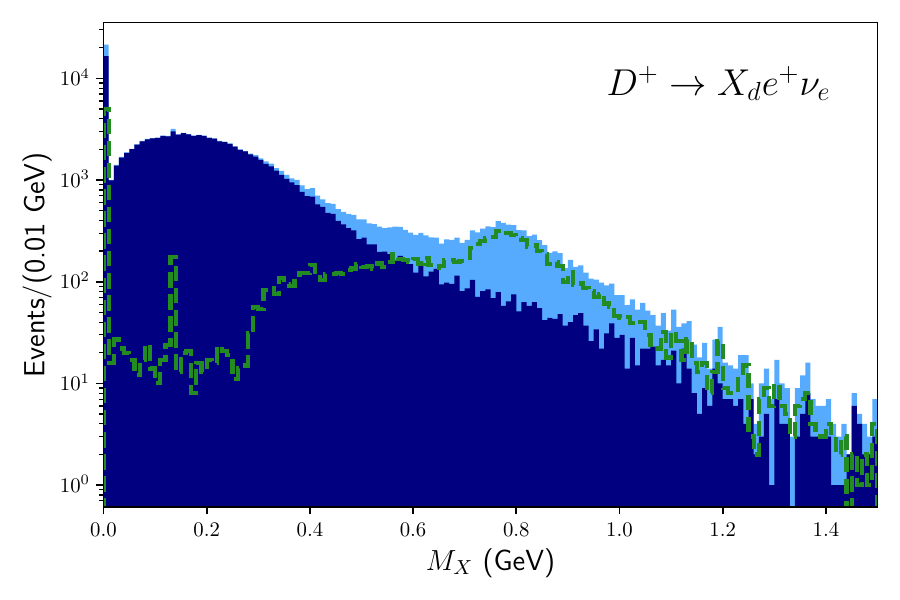}
\end{tabular}
    \caption{Estimated \mX distribution of (left) $c\to s$-tagged $D^+$ semileptonic decays and (right) $c\to d$-tagged $D^+$ semileptonic decays. In each case, the expected observed distribution is shown in light blue, the contribution of events with $K^0_L$ mesons and $K_S^0\to\pi^0\pi^0$ decays in the final state is shown in dark blue, and the true distribution of events passing selections shown as the dashed green histogram.}
    \label{fig:strangeness_tagging}
\end{figure}

The differentiation of inclusive decays proceeding through $c\to s$ and $c\to d$ transitions can be achieved by identifying the total strangeness of the hadronic system $X$. This can be determined from the number of observed $K^+$ mesons, $K_S^0\to\pi^+\pi^-$ decays, and $\Lambda\to p\pi^-$ decays with high efficiency. It should be noted again that the requirements on \Empm suppress contributions from decays with $K_L^0$ mesons, other $K_S^0$ decays, or neutrons in the final state, but that this can be corrected based on the observed decays with reconstructed $K_S^0\to\pi^+\pi^-$ and proton candidates.  Estimates of the capability of $c\to s$-tagging is shown in Fig.~\ref{fig:strangeness_tagging}. The residual contribution of events where the strangeness of the final state is misidentified (responsible for the large contribution of misidentified events in the $c\to d$ distribution) could be determined through a simultaneous template fit of the \mX distributions for the $c\to s$-tagged sample and the $c\to d$-tagged sample.

\subsection{Systematic uncertainties}

The size of systematic uncertainties in BESIII measurements of the differential spectra due to detector modeling are estimated by varying the assumed reconstruction resolution and the reconstruction and particle identification efficiencies within their stated uncertainties and repeating the procedure described in Sec.~\ref{sec:fastsim}. The calibration procedure described in Sec.~\ref{sec:calib} depends on the branching fractions that are assumed for the component exclusive branching fractions. Alternate spectra are produced based on the measured uncertainties of the exclusive branching fractions included in the fast simulation and checked to determine the magnitude of the dependence.

The dominating experimental systematic uncertainties are associated with the measurements of the exclusive branching fractions, which are all currently limited by BESIII sample sizes, and corrections for imperfect simulation of detector response, which are all also dominated by the size of data control samples used to estimate the correction. As such, we assume that the experimental systematic uncertainties of each of the measurements will scale with the square root of the collected sample's luminosity for data-taking scenarios possible in the near-future. We expect this assumption may not hold for samples larger than the currently collected $D^0$ and $D^+$ samples estimated in Table~\ref{tab:BESIIILumi} or samples ten times larger than the listed $D^+_s$ and $\Lambda_c$ samples that could be collected at a proposed Super-Tau-Charm Factory \cite{STCF}.

\subsection{Background contributions}

While we consider all foreseen detector effects on reconstructing and identifying signal events in the fast simulation, we do not consider possible contributions from backgrounds. Backgrounds will arise from the tag-side hadron being misidentified, the lepton candidate being a misidentified hadron, and contributions from $e^+e^-\gamma$ Dalitz decays of light mesons produced in the final state of charm hadron decays. The first category can be handled in a similar fashion to the inclusive semileptonic beauty measurements at Belle and Belle II~\cite {BelleQ2,BelleIIQ2}. The second category can be handled through a weighting procedure, where particle identification observables are used to weight the observed sample. The third and final category can be corrected for with a similar procedure to previous inclusive semileptonic charm measurements \cite{CLEO:2009uah,BESIIIDsIncSL,BESIII:2022cmg}, where these backgrounds are corrected for with data control samples in which the identified lepton has opposite charge to the expected flavour of the parent hadron. The expected impact of background subtraction on the statistical precision are estimated from the purities observed in the previous inclusive semileptonic charm measurements \cite{CLEO:2009uah,BESIIIDsIncSL,BESIII:2022cmg}, described in more detail in Sec.~\ref{sec:extraction}.

\section{Calibration procedure and sensitivity of raw spectral moments}
\label{sec:calib}

\begin{figure}
    \centering
    \includegraphics[width=0.7\linewidth]{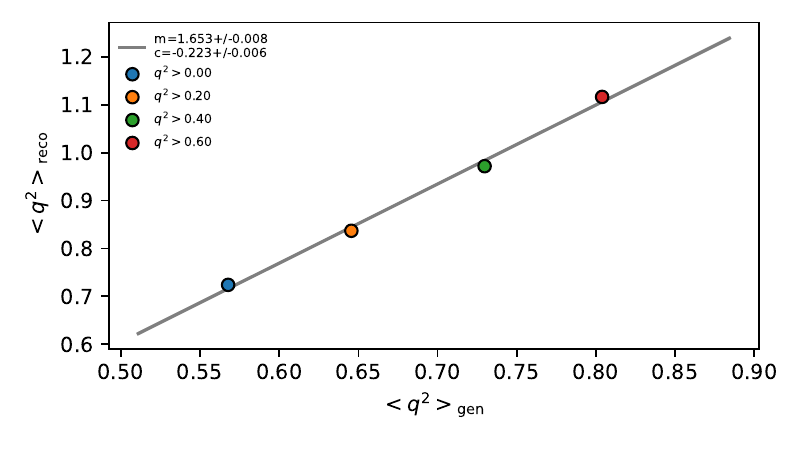}
    \caption{The linear calibration function determined for the first $q^2$ moment of the $D^0$ decay which is used to calibrate the reconstructed moments to the corresponding generator moments.}
    \label{fig:D0_q2_calib}
\end{figure}

\begin{figure}
    \centering
    \includegraphics[width=1.0\linewidth]{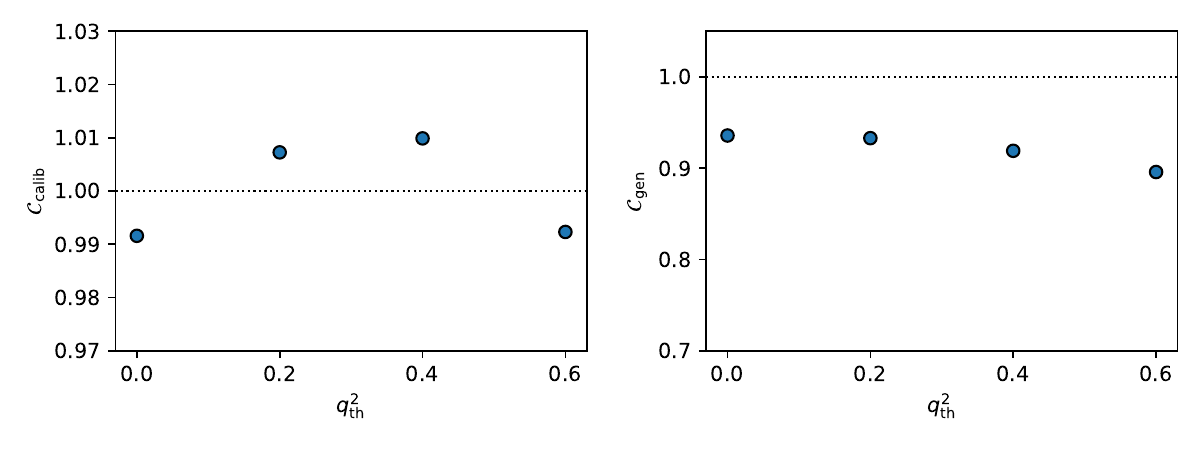}
    \caption{The $\mathcal{C}_\mathrm{calib}$ (left) and $\mathcal{C}_\mathrm{gen}$ (right) calibration factors for the $D^0$ decay. The former corrects for deviations from the linear calibration function, and the latter corrects for selection and acceptance effects.}
    \label{fig:D0_q2_C}
\end{figure}

We use the simulated events to determine the raw moments of $q^2$ and $E_\ell$ of order $n$ and for a progression of threshold cuts on $q^2$.
This is done in a three-step procedure following the approach from Ref.~\cite{BelleQ2,BelleIIQ2}. First, we apply a linear calibration as a function of $q^{2n}$ or $E_\ell^n$, exploiting the fact that the reconstructed spectral moments of the distributions of interest are mostly linearly shifted with respect to their true values. We determine the slope $c_n$ and intercept $m_n$ of the linear calibration function via a fit to the simulated samples that solves 
\begin{align}
 q^{2n}_{\text{cal}} & = \left(q^{2n}_{\text{reco}}  - c_n\right)/m_n   \, , & E_{\ell, \text{cal}}^{n} = \left(E_{\ell, \text{reco}}^n  - c_n \right) / m_n \, .
\end{align}
Here the subscript `reco' indicates the reconstructed moments and `cal' indicates the calibrated moments. We further derive individual calibration constants for $q^{2n}$ and $E_\ell^n$ and for each order $n$. In Figure~\ref{fig:D0_q2_calib} we show the linear calibration function for the first $q^2$ moment for the $D^0$ decay: the reconstruted and generated values are approximately linearly shifted.
Second, we correct the calibrated moments for deviations from the linear calibration function via a correction factor derived from simulation, 
\begin{align}
  \mathcal{C}_{\mathrm{calib}} & = \langle q^{2n}_{\mathrm{gen,sel}} \rangle   /  \langle q^{2n}_{\mathrm{cal}} \rangle \qquad \text{or} \qquad \langle E^{n}_{\ell,\mathrm{gen,sel}} \rangle   /  \langle E^{n}_{\ell,\mathrm{cal}} \rangle \qquad,
\end{align}
where the subscript \textit{`gen,sel'} indicates the generator level moments with selections applied. Third, we correct for the selection and acceptance effect with a correction factor, also derived from simulation,
\begin{align}
  \mathcal{C}_{\mathrm{gen}} & = \langle q^{2n}_{\mathrm{gen}} \rangle   /  \langle q^{2n}_{\mathrm{gen,sel}} \rangle \qquad \text{or} \qquad \langle E^{n}_{\ell,\mathrm{gen}} \rangle   /  \langle E^{2n}_{\ell,\mathrm{gen,sel}} \rangle \qquad 
\end{align}
where the subscript `gen' indicates the generator level moments without any selections or acceptance effects imposed. Both $\mathcal{C}_\mathrm{calib}$ and $\mathcal{C}_\mathrm{gen}$ are determined for each $q^2$ threshold and shown in Figure~\ref{fig:D0_q2_C}. The $\mathcal{C}_\mathrm{calib}$ factors range between 0.97 and 1.03 for the $\langle q^{2n} \rangle$ and are very close to unity for the $\langle E_\ell^n \rangle$. The $\mathcal{C}_\mathrm{gen}$ factors range between 0.75 and 1.00 for both $\langle q^{2n} \rangle$ and $\langle E_\ell^n \rangle$. The origin of these large factors are decays that are only partially reconstructed due to the presence of e.g. $K_L$, and are rejected by the $|E_\mathrm{miss} - p_\mathrm{miss}| > 0.5\,\mathrm{GeV}$ selection (cf. Fig.~\ref{fig:DpEmPm}). In Figure~\ref{fig:D0_q2_C} we show the calibration factors for the first $q^2$ moment for the $D^0$ decay at different thresholds $q^2_\mathrm{th}$.
Both $\mathcal{C}_\mathrm{calib}$ and $\mathcal{C}_\mathrm{gen}$ exhibit a similar behaviour among the four decays considered. A more detailed discussion on the linear calibration functions and the multiplicative calibration factors can be found in Appendix~\ref{app:calibration}. 

\begin{figure}
    \centering
    \includegraphics[width=\linewidth]{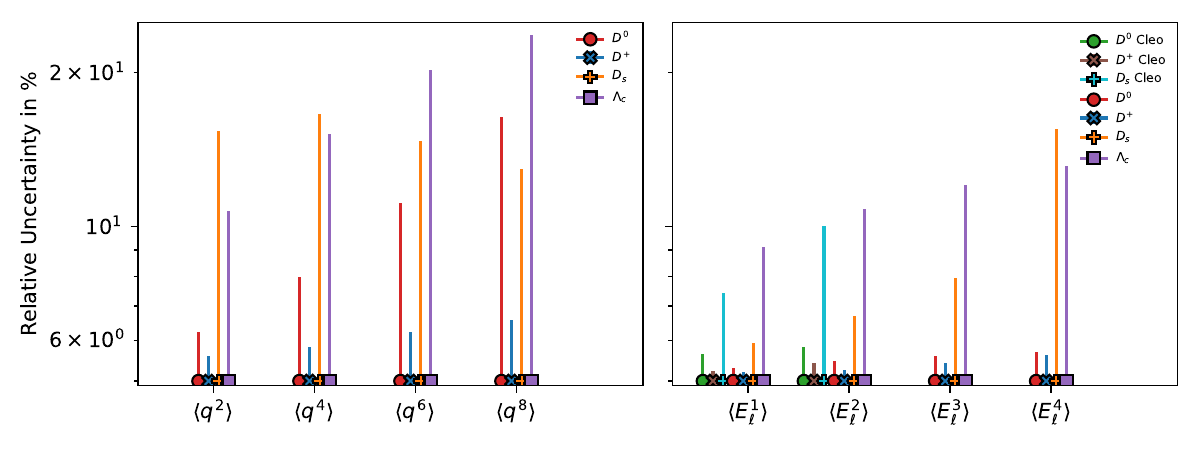}
    \caption{Relative uncertainties in \% on the determined moments with the calibration procedure described in the text. The uncertainties include statistical and systematic uncertainties. We only show the uncertainty for the moments with $q^2_\mathrm{th} = 0$. We also show relative uncertainty on the moments determined by CLEO from Ref.~\cite{CLEO:2009uah,Gambino:2010jz}.}
    \label{fig:uncertainty_moments}
\end{figure}

Applying all outlined calibration steps, we determine the spectral moments via
\begin{align}
 \langle q^{2n} \rangle & = \sum_i  q^{2n}_{i \, \text{cal}} \times \mathcal{C}_{\mathrm{calib}} \times \mathcal{C}_{\mathrm{gen}} \, ,  &   \langle E_\ell^{n} \rangle & = \sum_i  E^{n}_{\ell, i, \text{cal}} \times \mathcal{C}_{\mathrm{calib}} \times \mathcal{C}_{\mathrm{gen}} \, , 
\end{align}
with $i$ denoting a given event. The absolute precision, including both statistical and systematic uncertainties, that can be achieved for the moments based on our simulation outlined above is listed in Table~\ref{tab:uncertainty_moments} and shown in Figure~\ref{fig:uncertainty_moments}.

\begin{table}[]
    \centering
    \caption{Relative uncertainties in \% on the determined moments with the calibration procedure described in the text. The uncertainties include statistical and systematic uncertainties. We only show the uncertainty for the moments with $q^2_\mathrm{th} = 0$.}
    \label{tab:uncertainty_moments}
    \vspace{2ex}
    \input{tables/moments}
\end{table}

\section{The Heavy Quark Expansion for charm}\label{sec:theory}
\subsection{Theoretical framework}
The setup of the Operator Product Expansion (OPE) for inclusive $D\to X \ell \bar \nu_\ell$ decays differs from the standard, well-established setup for $B\to X_c \ell  \bar \nu_\ell$ decays due to the hierarchy between the charm-quark and light-quark decay products. In the latter case, the ratio $m_c/m_b$ is fixed when setting up the OPE. However, for $D \to X_s$ and $D \to X_d$ decays, the $s$ and $d$ quarks need to be integrated out at their respective scales. This introduces four-quark operators (weak annihilation) (see e.g. \cite{Bigi:2009ym,Breidenbach:2008ua} for a detailed discussion). As such, the OPE used to describe charm decays resembles more that of the $B\to X_u\ell \bar \nu_\ell$ case. The setup of the HQE is such that the hadronic parameters depend on the mass of the initial state meson. Therefore, the hadronic matrix elements differ between charm and beauty decays, but also between $D_s$ and $D$ decays.

From the theoretical side, extracting the HQE element from data requires the OPE expressions for the spectral moments at certain order in the $\Lambda_{\rm QCD}/m_c$ and $\alpha_s(m_c)$ expansions. Here we follow the approach of Ref.~\cite{Fael:2019umf}, which set up the HQE as an expansion in both $\Lambda_{\rm QCD}/m_c$, $\alpha_s(m_c)$ and $m_s/m_c$ for the $c\to s$ transition. The $c\to d$ expressions can be obtained by taking $m_s\to 0$. 

In the following, we only work up to order $1/m_c^3$, although the $1/m_c^4$ terms are also known \cite{Fael:2019umf} and could be included in a future analysis. We define the hadronic matrix elements $X$ for the two-quark operators following \cite{Fael:2019umf,Fael:2018vsp,Mannel:2018mqv}
\begin{align}
    2m_D X & \equiv \langle D| O^{2q}_X |D \rangle \ ,
\end{align}
where equivalent expressions for $D_s$ or $\Lambda_c$ can be written. The matrix elements are then\footnote{These are defined in the reparametrization invariant (RPI) basis. See \cite{Fael:2018vsp} for a conversion to the spatial derivative basis.}
\begin{align}
 2m_D \mu_\pi^2 & =  \langle \bar{c}_v (iD^2) c_v \rangle\\
 2m_D \mu_G^2 & =  \langle \bar{c}_v (iD_\alpha) (iD_\beta) (-i\sigma^{\alpha\beta}) c_v \rangle
 \end{align}
 and at order $1/m_c^3$
 \begin{align}
2m_D \rho_{LS}^3  & =  \frac{1}{2} \langle \bar{c}_v \left\{ (iD_\alpha),\left[ivD, iD_\beta\right]\right\} (-i\sigma^{\alpha\beta})c_v \rangle  \\
  2m_D \rho_D^3 & =  \frac{1}{2} \langle \bar{c}_v \left[ (iD_\mu),\left[ivD, iD^\mu\right]\right]c_v \rangle 
\end{align}
where we have omitted the initial state $D$ mesons for simplicity but note that the matrix elements depend on the initial state meson. In addition, unlike in the setup for $B\to X_c$ inclusive decays, the four-quark (weak annihilation) operators remain in the OPE. Following Ref.~\cite{Fael:2019umf}, we define
\begin{equation}
  2 m_D T_i(\mu_{\rm WA}) \equiv \langle D| O^{4q}_i |D \rangle , \quad \mbox{with } i=1,2.
\end{equation}
where $\mu_{\rm WA}$ is the weak annihilation scale. At order $1/m_c^3$ only two operators contribute
\begin{align}\label{eq:ops}
O_1 &=  (\bar{c}_v  \slashed v P_L s) \, (\bar{s} \slashed v P_L c_v), \notag \\
O_2 &=  (\bar{c}_v  \gamma^\mu P_L s) \, (\bar{s} \gamma_\mu P_L c_v). 
\end{align}
The weak annihilation operators $T_1$ and $T_2$ mix under renormalization with the Darwin term \cite{Breidenbach:2008ua,Bigi:2009ym}. This introduces only one new operator \cite{Fael:2019umf}
\begin{equation}
  \tau_0 = 128 \pi^2 \left(T_1-T_2  \right) +8\log \left( \frac{\mu_{\rm WA}^2}{m_c^2} \right)
   \rho_D^3 .
\end{equation}
We note that our setup, which follows \cite{Fael:2019umf}, differs from \cite{Gambino:2010jz, Ligeti:2010vd,Wu:2024jyf} due to the three expansion parameters that we employ and through the definition of the four-quark operators given above. In the latter, the WA matrix elements are named $B_{\rm WA}$. Specifically, our expressions differ in the non-logarithmic terms from those of $b\to c$ due to a difference in the matching of the two-quark operators for the $c\to s$ transition. This subtle point was discussed in \cite{Fael:2019umf} and overlooked in \cite{Gambino:2010jz, Ligeti:2010vd, Wu:2024jyf}.  In addition, converting $\tau_0$ to $B_{\rm WA}$ also requires setting the scale $\mu_{\rm WA}$, which in \cite{Gambino:2010jz} was set at $\mu_{\rm WA}=0.8$ GeV. 

The operators $T_1$ and $T_2$ discussed above are non-valence Weak Annihilation (WA) operators, which do not depend on the flavour of the spectator quark. For $D^0$ decays, only non-valence WA contributes, however for $D_s$ and $D^+$ decays the valence WA also enters weighted by the relevant CKM factors (see expressions in Sec.~6 of \cite{Fael:2019umf}). These operators are defined through \eqref{eq:ops} by replacing $s\to q$, where $q$ is the valence spectator quark. The non-valence WA contribution is similar for $D^0$ and $D^+$ but differs for $D_s$ decays due to $SU(3)$ breaking effects that also affect the other HQE elements. The valence and non-valence WA could be disentangled by measuring the WA parameters, $\tau_0$, from $D^0, D^+$ and $D_s$ decays \cite{Bigi:2009ym}. It is important to extract these WA parameters from inclusive charm decays as they form a crucial input into the $D^0, D^+$ and $D_s$ lifetime calculations.

The WA operators have been obtained from the total rates in \cite{Ligeti:2010vd}. In \cite{Gambino:2010jz}, the CLEO inclusive charm data were converted to lepton energy moments and used to extract both the non-valence and valence WA operators. As discussed above, their setup differs from \cite{Fael:2019umf} employed here, making it more challenging to disentangle the effect of the logarithmic $\rho_D^3$ terms and the WA effects. Moreover, $q^2$ moments are currently not available. As such, redoing the extraction of WA combined with the first extraction of the other HQE parameters in charm would be beneficial. There is also quite some progress in determining the HQE elements directly from lattice QCD calculations \cite{Gambino:2017vkx, Gambino:2022dvu, Nefediev:2024mjk, Kellermann:2023yec}. In the future, it will be intriguing to compare extractions from data with the lattice calculations.

In order to extract the HQE parameters, we define the moments of the spectrum as
\begin{equation} 
 \mathcal{M}_n  \equiv \frac{1}{\Gamma_0} \int   (M)^n \frac{\text{d} \Gamma}{\text{d} M} \, \text{d} M \ .
\end{equation}
where $M$ is the lepton energy $E_\ell$ or the dilepton invariant mass $q^2$ and $\Gamma_0$ is the tree-level normalization factor. In the current work, we only consider lepton energy and $q^2$ moments. The latter have the advantage that they are reparametrization invariant (RPI) observables that depend on a reduced set of observables. Specifically, $\rho^3_{LS}$ and $\mu_\pi^2$ do not enter the $q^2$ moments \cite{Mannel:2018mqv,Fael:2019umf}. The integration limits are over the allowed phase space depending on the employed kinematic thresholds on the lepton energy or $q^2$. Here we work with normalized moments defined as
\begin{equation}\label{eq:momdef}
\langle M^n\rangle \equiv \frac{\mathcal{M}_n}{\mathcal{M}_0} \ .
\end{equation}
These moments can be converted to centralized moments, which are customary to use $B\to X_c \ell \bar \nu_\ell$ analyses (see e.g.~\cite{Finauri:2023kte,Bernlochner:2022ucr}). For our exploratory study, we however fit the raw moments as defined in \eqref{eq:momdef}.
The expressions for the lepton-energy and $q^2$ moments at tree level are given in \cite{Fael:2019umf}. For the $\Lambda_c$ decays, the expressions are similar to those for the $D$ decays, except that all $\sigma\cdot G$ terms cancel due to the spin structure and thus $\mu_G^2$ and $\rho_{LS}^3$ do not contribute\footnote{Technically, we first have to split off the spatial and time derivatives in $\mu_G^2$ by converting from the RPI basis as $\mu_G^2 \to (\mu_G^2)^\perp -\rho_D^3/m_c -\rho_{LS}^3/m_c$. We can then set $(\mu_G^2)^\perp$ and $\rho_{LS}^3$ to zero.}. In the following, we consider moments up to $n=4$. We note that especially these higher order moments are very sensitive to the HQE parameters and would suffer from larger theoretical uncertainties due to neglected $1/m_c^4$ HQE parameters.

Finally, using information on the total decay semileptonic width allows for an extraction of $|V_{cs(d)}|$. At leading order in $\alpha_s$, we have 
\begin{align}
    \frac{\Gamma(D \to X_s \ell  \bar\nu_\ell)}{\Gamma_0}  &= 
  \left( 1-8 \rho - 10 \rho^2 \right) \left(1-\frac{\mu_\pi^2 - \mu_G^2}{2m_c^2}\right)
  +\left(-2 - 8 \rho \right) \frac{\mu_G^2}{m_c^2}
  +6 \frac{\rho_D^3}{m_c^3}   +\frac{\tau_0}{m_c^3}\, ,
\end{align}
where $\rho= (m_s/m_c)^2$ and $\Gamma_0= G_F^2 m_c^5 |V_{cs}|^2/(192\pi^3)$ and equivalent for $D\to X_d$ decays by putting $\rho\to 0$ and replacing $|V_{cs}|\to |V_{cd}|$. 

\subsection{Theory challenges}
For a precision determination of the HQE parameters in charm, the setup of the HQE for charm at leading order in $\alpha_s$ does not suffice due to the large value of the strong coupling constant at the relevant scale $m_c$. 
Practically this means higher-order $\alpha_s$ corrections should be included, while getting a reliable estimate for the missing higher-orders in $\alpha_s$ is challenging. Clearly, a simple scale variation (often done for $B$ decays) would render much larger uncertainties. For charm decays, $\alpha_s$ corrections for $q^2$ and lepton energy moments can be obtained from \cite{Fael:2022frj} which uses an expansion in $\delta = 1-\sqrt{\rho}$ with $\rho \equiv (m_q/m_Q)^2$ for the $m_Q\to m_q$ transition. For massless quarks, the $\alpha_s$ corrections can be obtained from \cite{Chen:2022wit} for $q^2$ moments. For the lepton energy moments in charm decays, \cite{Gambino:2010jz} implemented the $\alpha_s^2$ BLM corrections obtained from \cite{Aquila:2005hq} for different lepton energy threshold cuts. For the total rate, $\alpha_s^2$ corrections are available \cite{Melnikov:2008qs}. To our knowledge, for $q^2$ moments, no dedicated calculations for $c\to s$ transitions are available. They could be obtained by adapting existing calculations for inclusive $b\to c$ semileptonic transitions. For the latter, the full $\alpha_s^2$ corrections to the $q^2$ moments were recently obtained \cite{Fael:2024gyw}\footnote{Slightly before that the BLM corrections for the $q^2$ moments were calculated \cite{Finauri:2023kte}}. For the lepton energy moments, non-BLM corrections are only known for $B$ decays for fixed $m_c/m_b$ masses and for several kinematical cuts \cite{Biswas:2009rb}. In addition, for $b\to c$ decays even the $\alpha_s^3$ corrections to the total rate and kinematic moments without any kinematic cut are available \cite{Fael:2020tow, Fael:2022frj}. 

In addition to the NNLO (or higher) corrections, a short-distance mass scheme for the charm quark is required to avoid the bad perturbative behavior of the pole mass. In $B\to X_c$ decays it is customary to use the kinetic mass \cite{Bigi:1994ga,Bigi:1996si}, which relies on a cut of $\Lambda_{\rm QCD}<\mu<m_Q$. See also Refs.~\cite{Bauer:2004ve,Bauer:2002sh} for a discussion on different mass schemes in $B$ decays. For charm decays, this makes the allowed window for $\mu$ rather small. Refs.~\cite{Gambino:2010jz,Wu:2024jyf} use the kinetic scheme for charm with $\mu=0.5$ GeV, however, as discussed in \cite{Fael:2018vsp} (see also \cite{Ligeti:2010vd}), it remains to be seen if $\mu\simeq0.5$ GeV is sufficiently away from $\Lambda_{\rm QCD}$. Alternatively, one could employ the $\overline{\rm MS}$ scheme at $\mu=m_Q$ \cite{Czarnecki:1994pu}, although it has been argued that this scheme is less suitable for decays of the $b$ and $c$ quarks given that the typical energy release is much lower than $m_Q$ \cite{Melnikov:2000qh}. In Ref.~\cite{Ligeti:2010vd}, the perturbative series of the $1S$ \cite{Hoang:1998ng,Hoang:2005zw} and PS scheme \cite{Beneke:1998rk}, finding a reasonably well behaved series for the former. 

In order to set up a full charm semileptonic program, a suitable charm mass scheme should be investigated further. This also applies to the HQE parameters themselves. In the $B\to X_c \ell \bar\nu_\ell$ analysis, the kinetic scheme for the HQE parameters is also employed. This kinetic scheme change introduces additional perturbative corrections that multiply the coefficients of $\mu_\pi^2$ and $\rho_D^3$. The effect on the four-quark parameters $\tau_0$ remains to be investigated. In addition, this kinetic scheme change of the HQE parameters suffers from the same issue as the kinetic charm  mass definition. 

As such, the dedicated experimental program in inclusive charm proposed here should be combined with detailed theoretical study of the perturbative corrections to inclusive charm decays.

\section{Towards an extraction of the Darwin and other HQE operators}
\label{sec:extraction}
To check the future potential of an inclusive charm program at BESIII, we perform an exploratory study and analyze the $q^2$ and $E_\ell$ moments. Given the theoretical challenges described above, we consider only the achievable experimental uncertainty on the extracted HQE parameters. Our sample data contains $\{D^0, D^+, D_s, \Lambda_c\} \to (X_s + X_d) \ell \bar{\nu}$ decays. As discussed, in the future these samples could be separated as is done for semileptonic $B\to X_c$ versus $B\to X_u$ decays. For this first study, we neglect the $c\to d$ contribution in the theoretical expressions for simplicity. 
In addition, we correct for the lepton energy acceptance when applying the $\mathcal{C}_\mathrm{gen}$ factor, so that the measured moments have only the quoted $q^2_{th}$ threshold selections applied.

Specifically, we work up to order $1/m_c^3$ in the HQE and only include $\alpha_s$ corrections. It is customary to include an additional theoretical uncertainty due to excluded higher orders in the HQE (and $\alpha_s$ expansion) (see e.g. the discussion in \cite{Finauri:2023kte}). We do not include any theory uncertainties in the fit, as the aim of our work is to determine the experimental feasibility of such measurements. Estimates of the theory uncertainties requires a dedicated program addressing the challenges detailed above We also do not consider any external inputs for the $\mu_G^2$ parameter which can in principle be linked to the mass difference of the initial state and its first excited state.

For the strange quark mass, we use the $2+1+1$ Lattice QCD FLAG averages \cite{FlavourLatticeAveragingGroup:2019iem,EuropeanTwistedMass:2014osg,FermilabLattice:2018est,Chakraborty:2014aca,Lytle:2018evc}
\begin{equation}
\overline{m}_s(2\; {\rm GeV}) = (93.44 \pm 0.68)\; {\rm MeV} \,.
\end{equation}

As discussed, the choice of the charm-quark mass definition is more challenging. Since our aim is to get an estimate for the attainable experimental precision only, we do not discuss in detail the different mass schemes for the quarks. For our study, we use the $\overline{\rm MS}$ definition, obtained from $2+1+1$ Lattice QCD FLAG average
\cite{FlavourLatticeAveragingGroup:2019iem,EuropeanTwistedMass:2014osg,FermilabLattice:2018est,Chakraborty:2014aca,Lytle:2018evc,Alexandrou:2014sha}
\begin{equation}
    \overline{m}_c(\overline{m}_c) =  (1.280 \pm 0.013) \;{\rm GeV} \ .
\end{equation}
We treat both masses as external inputs and fixed parameters. 
We note that this choice gives rather large $\alpha_s$ corrections, although the convergence of the series can only be checked by including higher order terms in $\alpha_s$. Comparing to the kinetic scheme with $\mu=0.5\,\mathrm{GeV}$  and $m_c^{\rm kin}=1.4\,\mathrm{GeV}$, we find that the $\alpha_s$ contributions for the lepton energy moments are smaller for the kinetic scheme. Its contribution is further reduced by including also the scheme change of the HQE parameters (as done in \cite{Gambino:2010jz}). For the $q^2$ moments, we however observe large $\alpha_s$ corrections in the kinetic scheme when including the perturbative contribution to $\rho_D^3$, which should be investigated further.
Alternatively, $m_c$ and $m_s/m_c$ could be obtained from the fit (see discussion for $B\to X_c$ decays in \cite{Gambino:2013rza}). 

For the coupling constant, we use 
\begin{equation}\label{eq:as}
    \alpha_s(\overline{m}_c) = 0.386  \,,
\end{equation}
obtained from RunDec with $n_f=3$ active flavours \cite{Chetyrkin:2000yt,Schmidt:2012az,Herren:2017osy}. 
Recently, \cite{Wu:2024jyf} used the inclusive charm branching ratios together with input for the HQE parameters from the $B\to X_c \ell \bar \nu_\ell$ decays to extract $\alpha_s$ employing the kinetic scheme for the charm mass. Our input value is in good agreement with the value extracted in that way. In a similar way as in \cite{Wu:2024jyf}, a future moments analysis could also allow the determination of the value of the strong coupling constant directly from the data. 

The fit is performed in the following way: We define the $\chi^2$ function as 
\begin{equation}
    \chi^2 = \chi^2(\mu_\pi^2, \mu_G^2, \rho_D^3, \rho_{LS}^3, \tau_0)´\,,
\end{equation}
depending only on the HQE parameters up to $\mathcal{O}(1/m_b^3)$ and the weak annihilation coefficient $\tau_0$. 
The experimental data used in the fit are the fully inclusive ($q^2_{\rm th} = 0 \, \text{GeV}^2$) $\langle q^{2n} \rangle$ and $\langle E_\ell^n \rangle$ moments with $n = \{1, 2, 3, 4\}$. We take into account the full statistical and systematic covariance between the two sets of spectral moments and the different orders. We perform the fit to the simulated data, construct an Asimov data set~\cite{Cowan:2010js} based on the fit result, and refit the Asimov data to determine the experimental precision. 
To take into account backgrounds from wrongly reconstructed tags, the statistical uncertainty is scaled by $1/\sqrt{P}$, where $P$ is the estimated purity based on the previous BESIII inclusive semileptonic analyses \cite{CLEO:2009uah,BESIIIDsIncSL,BESIII:2022cmg}.

\begin{figure}
    \centering
    \includegraphics[width=1\linewidth]{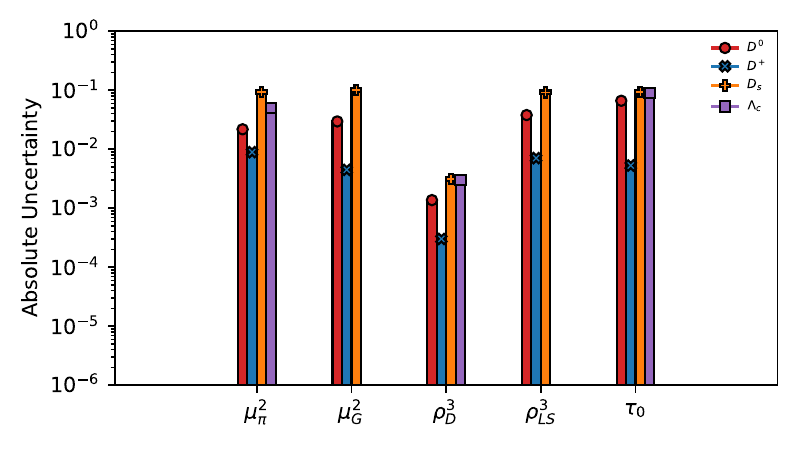}
    \caption{Estimated precision on the HQE and weak annihilation parameters in the four decays considered from the simulation described in the text. The error bars include all systematic uncertainties described in the text.}
    \label{fig:HQE_parameters}
\end{figure}

The resulting estimates on the attainable absolute uncertainty on the extracted HQE parameters are given in Table~\ref{tab:HQE_parameters} and shown in Figure~\ref{fig:HQE_parameters}. We obtain similar uncertainties when using $m_c$ in the kinetic scheme. These estimates show an excellent experimental precision and highlight the feasibility of the suggested program at BESIII. In particular, the experimental uncertainty on the $\rho_D^3$ extraction looks very promising. The precision on $\mu_G^2$ may be further improved by putting an external constraint from the mass splitting. As discussed, the uncertainty attainable on the weak annihilation parameter $\tau_0$ is challenging to compare with the previous extraction in \cite{Gambino:2010jz} due the mixing with $\rho_D^3$. A simple conversion based on the phase-space factors allows an estimate of the improvement. We find over a factor of two up to an order of magnitude improvement. In \cite{Gambino:2010jz}, the WA parameters were found to be consistent with zero. It will be important to determine if they remain consistent with zero even with improved precision. For the $\Lambda_c$ decays, no HQE parameters have been extracted before. For $\rho_{LS}^3$, which is only constrained through the lepton energy moments, we note that this parameter has only a small prefactor. Due to RPI, it does not enter the $q^2$ moments nor the total rate.

Comparing with the relative uncertainty on the measured moments in Fig.~\ref{fig:uncertainty_moments}, we note that moments of the $D_s$ and $\Lambda_c$ modes are determined least precise. This is also reflected in the estimated precision on the extracted HQE parameters. We highlight the excellent prospects for the extractions from $D^+$ decays due to its large branching fraction. We note that the inclusion of $\alpha_s^2$ corrections or a different mass definition would shift the central values of the theoretical expressions and might also affect the absolute uncertainty on the extraction. A similar note holds for including a theoretical uncertainty, which allows more flexibility for the fit to accommodate the measured moments. As such, the exact values of the estimated uncertainties should be taken with caution and mainly identify points of improvement for a future experimental analysis.

\begin{figure}
    \centering
    \includegraphics[width=1\linewidth]{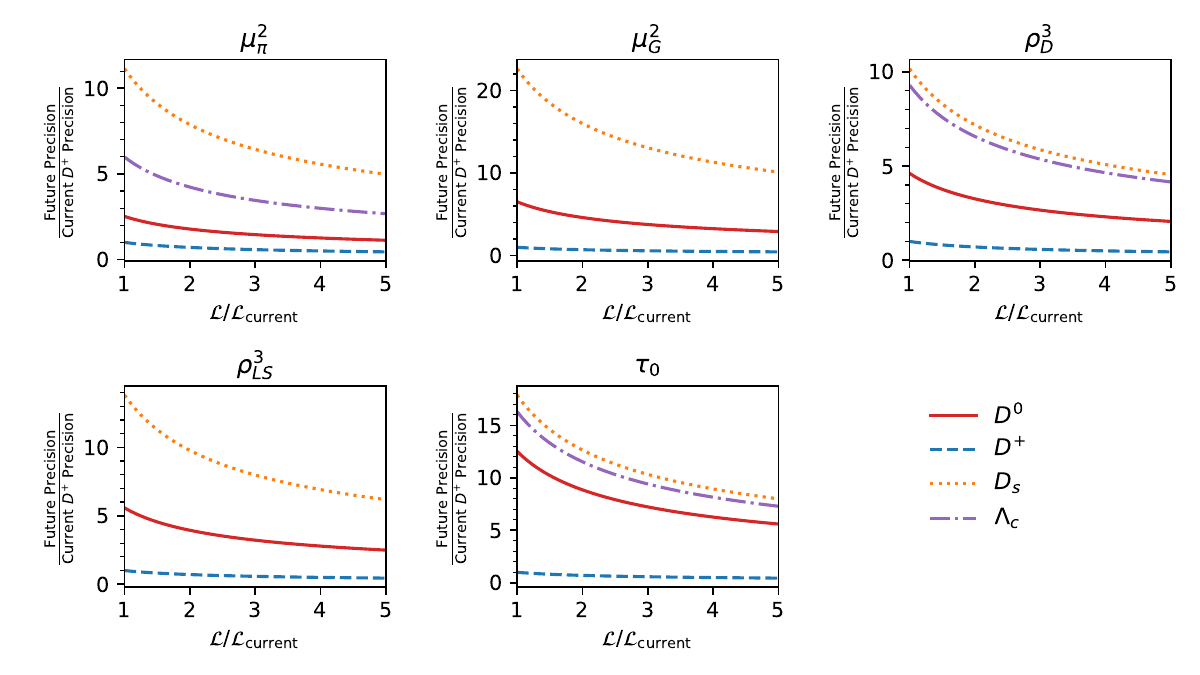}
    \caption{Estimated precision on the HQE and weak annihilation parameters with increased integrated luminosity relative to the current precision for $D^+$.  We assume that all uncertainties considered scale with $1/\sqrt{\mathcal{L}}$ in the given range. 
 }\label{fig:HQE_parameters_scaling}
\end{figure}

The uncertainty budget of the $D^0$, $D^+$, $D_s$, and $\Lambda_c$ is given in Tables~\ref{tab:D0_systematics}, \ref{tab:Dp_systematics}, \ref{tab:Ds_systematics}, and \ref{tab:Lc_systematics}, respectively. 
In each decay mode, the statistical uncertainty can be improved with a larger BESIII data set, and the calibration uncertainty (labeled as `MC stat.') can be improved with increases in the integrated luminosity of the simulated data set. The track reconstruction efficiency is also a leading uncertainty. We stress that the  $D_s$ and $\Lambda_c$ modes specifically would benefit from more precise measurements of the exclusive branching ratios.  We show the evolution of the precision estimates in Figure~\ref{fig:HQE_parameters_scaling}, where we assume that all uncertainties will scale with the square root of the integrated luminosities and that BESIII will keep 40 times the integrated luminosity for their simulation with respect to the recorded data. Increased samples of $D_s$ mesons and $\Lambda_c$ baryons at BESIII would allow for competitive precision on the HQE parameters for all four flavours of charm hadrons.

\begin{table}[]
    \centering
    \caption{Estimated absolute precision $\times 10^3$ on the HQE and weak annihilation parameters in the four decays considered from the simulation described in the text. The quoted uncertainty includes all systematic uncertainties considered, and a more detailed separation into individual sources of uncertainties can be found in the text.}
    \label{tab:HQE_parameters}
    \vspace{2ex}
    \input{tables/HQE_parameters}
\end{table}

\begin{table}[]
    \centering
    \caption{Absolute uncertainties on the HQE and weak annihilation parameters for the $D^0$ channel. A description of the systematic uncertainties can be found in the main text. The values are multiplied by $\times 10^3$. 
       }
    \label{tab:D0_systematics}
    \vspace{2ex}
    \input{tables/D0_systematics}
\end{table}

\begin{table}[]
    \centering
    \caption{Absolute uncertainties on the HQE and weak annihilation parameters for the $D^+$ channel. A description of the systematic uncertainties can be found in the main text. The values are multiplied by $\times 10^3$.}
    \label{tab:Dp_systematics}
    \vspace{2ex}        
    \input{tables/Dp_systematics}
\end{table}

\begin{table}[]
    \centering
    \caption{Absolute uncertainties on the HQE and weak annihilation parameters for the $D_s$ channel. A description of the systematic uncertainties can be found in the main text. The values are multiplied by $\times 10^3$.}
    \label{tab:Ds_systematics}
    \vspace{2ex}    
    \input{tables/Ds_systematics}
\end{table}

\begin{table}[]
    \centering
    \caption{Absolute uncertainties on the HQE and weak annihilation parameters for the $\Lambda_c$ channel. A description of the systematic uncertainties can be found in the main text. The values are multiplied by $\times 10^3$.}
    \label{tab:Lc_systematics}
    \vspace{2ex}        
    \input{tables/Lc_systematics}
\end{table}

\FloatBarrier

\section{Outlook}\label{sec:outlook}
As demonstrated, a future program of the study of inclusive charm decays at BESIII is extremely promising. Such a program will improve measurements of the lepton energy moments and  allow for the first determinations of the reparametrization invariant $q^2$ moments of the kinematical spectrum of $D^0, D^+, D_s$ and $\Lambda_c$ inclusive decays. This opens the way for studying the HQE for the first time in baryonic decays and making a direct comparison between different initial state mesons. Our analysis shows that such a measurement program is achievable, and furthermore highlights how the measurements could further improve. The $D_s^+$ and $\Lambda_c$ measurements have particularly strong prospects for improvement with more data. 

We have performed an exploratory analysis of the estimated moments, obtaining a promising attainable precision on the $\{\mu_\pi^2, \mu_G^2, \rho_D^3, \rho_{LS}^3\}$ HQE elements and the important weak annihilation $\tau_0$ parameter. The precision on the extracted HQE parameters could be further improved by analyzing moments of the spectrum at different kinematic cuts on the $q^2$ or lepton energy, in a similar fashion to the $B\to X_c \ell\bar\nu_\ell$ analysis. In addition, moments of the hadronic invariant mass $M_X$ could be studied. For this work, we extracted normalized raw moments of the spectrum. In a future analysis, it would also be interesting to compare the theoretical or experimental benefits of normalized raw versus centralized moments.

Combining the extracted HQE elements with measurements of the inclusive branching ratio allows for the extraction of $|V_{cs}|$ and $|V_{cd}|$ for the first time from inclusive decays. The latter requires separating the $c\to s$ from $c\to d$ transition, which as outlined seems feasible in a future analysis. Using the determined HQE parameters, we obtain an experimental precision for $|V_{cs}|$ of 3.3\% for $D^0$ and $D^+$ (with a branching fraction uncertainty contribution of 1.1\%~\cite{CLEO:2009uah}), and 3.8\% for $D_s^+$ (1.3\% due to the branching fraction~\cite{BESIIIDsIncSL}). Given the long-standing puzzle between inclusive and exclusive $|V_{cb}|$, it would be charming to obtain inclusive $|V_{cs}|$ and $|V_{cd}|$ from data. 

The HQE parameters have been extracted with good precision from semileptonic $B\to X_c \ell \bar{\nu}$ decays, but have never been studied in inclusive charm decays. Extracting the HQE parameters with high precision requires improvements in theory. At minimum, the NNLO and possibly N3LO corrections should be included, and a consistent charm mass definition should be employed. This would then allow for the establishment of a theoretical framework analogous to that used in $B\to X_c \ell\bar \nu_\ell$ decays. In the future, this framework could be merged with the open-source package Kolya which provides theoretical expressions for inclusive $B$ decays \cite{Kolya, kolyapaper}. 

Comparing the HQE parameters among the different charm species will provide the first test of $SU(3)$ flavour symmetry within the HQE and allow valence and non-valence weak annihilation parameters to be separated. Whether or not a HQE analysis of the experimental moments converges will provide an additional and valuable experimental test of the validity of the HQE for charm. Moreover, comparing the HQE parameters between beauty and charm allows for a further test of non-perturbative QCD among different generations. In the infinite mass limit, the HQE parameters in beauty in charm are equal, but in the standard HQE approach they depend in a non-trivial way on the mass of the initial decaying hadron. In Refs.~\cite{Fael:2019umf, Gambino:2010jz}, the difference between beauty and charm elements is estimated. With the outlined program, these relations can be tested with data. 

In addition to these comparisons among different quark species, determining the HQE elements themselves is important as they enter the theoretical predictions for charm lifetimes. Furthermore, the weak annihilation parameters also enter $B\to X_{s,d}\ell\ell$ and $B\to X_u \ell\bar\nu_\ell$ predictions. 
In this way, the envisioned charm program reaches across different families and would allow for compatibility checks between hadron sectors within the Standard Model. There is a grandeur in this view of semileptonic decays.

\section*{Acknowledgements}
The work of KKV is
supported in part by the Dutch Research Council (NWO) as part of the project Solving
Beautiful Puzzles (VI.Vidi.223.083) of the research programme Vidi.
The work of FB is supported by DFG Emmy-Noether Grant No. BE 6075/1-1 and BMBF Grant No. 05H21PDKBA
The work of MP is supported by the German Research Foundation (DFG) Emmy-Noether Grant No. 526218088.
SM and GW are grateful for support from the UK Science and Technology Facilities Council.
\clearpage
\pagebreak
\bibliography{main.bib}

\appendix

\section{Calibration}
\label{app:calibration}

\begin{figure}
    \centering
    \includegraphics[width=\linewidth]{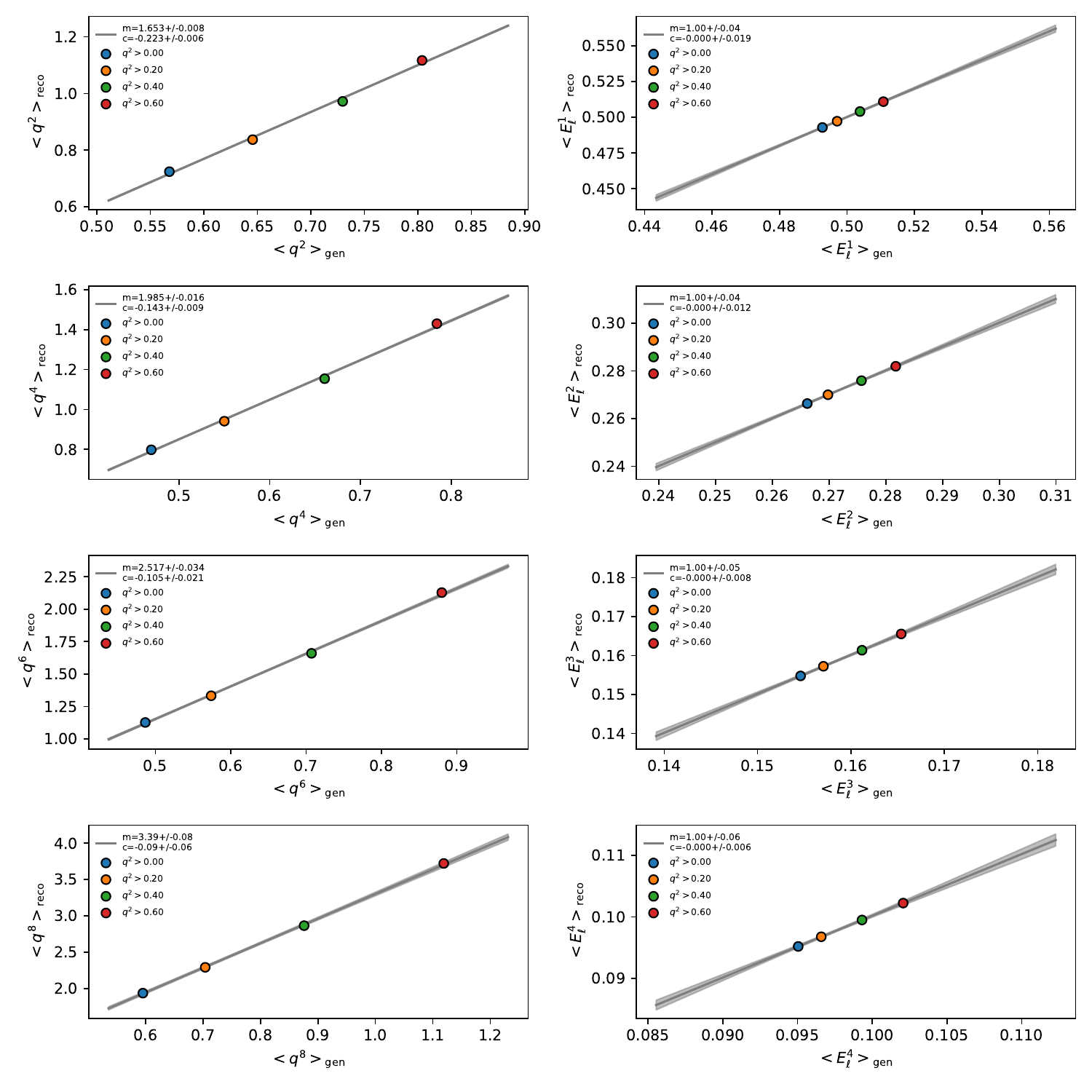}
    \caption{$D^0$ raw moment calibration}
    \label{fig:calibration_D0}
\end{figure}

\begin{figure}
    \centering
    \includegraphics[width=\linewidth]{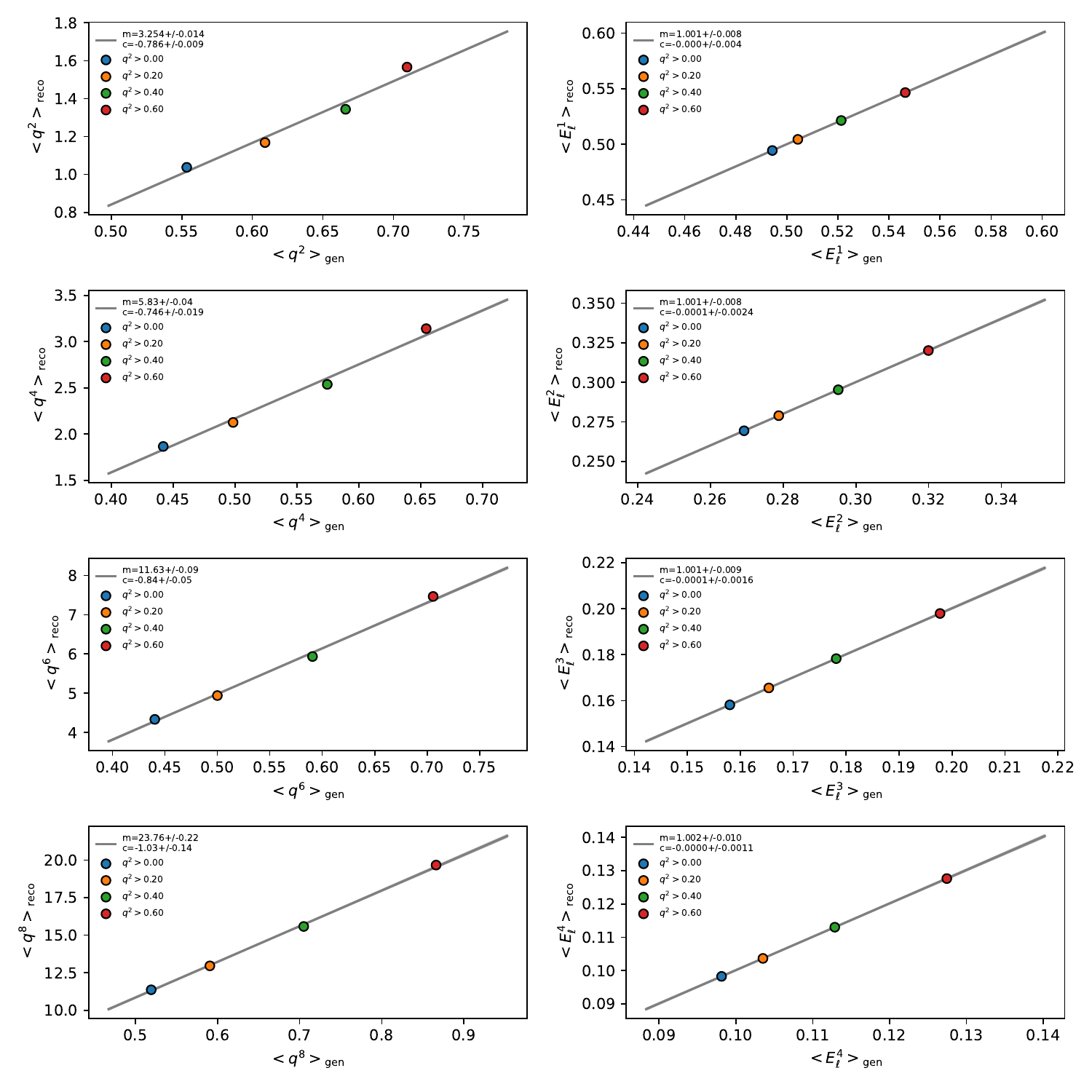}
    \caption{$D^+$ raw moment calibration}
    \label{fig:calibration_Dp}
\end{figure}

\begin{figure}
    \centering
    \includegraphics[width=\linewidth]{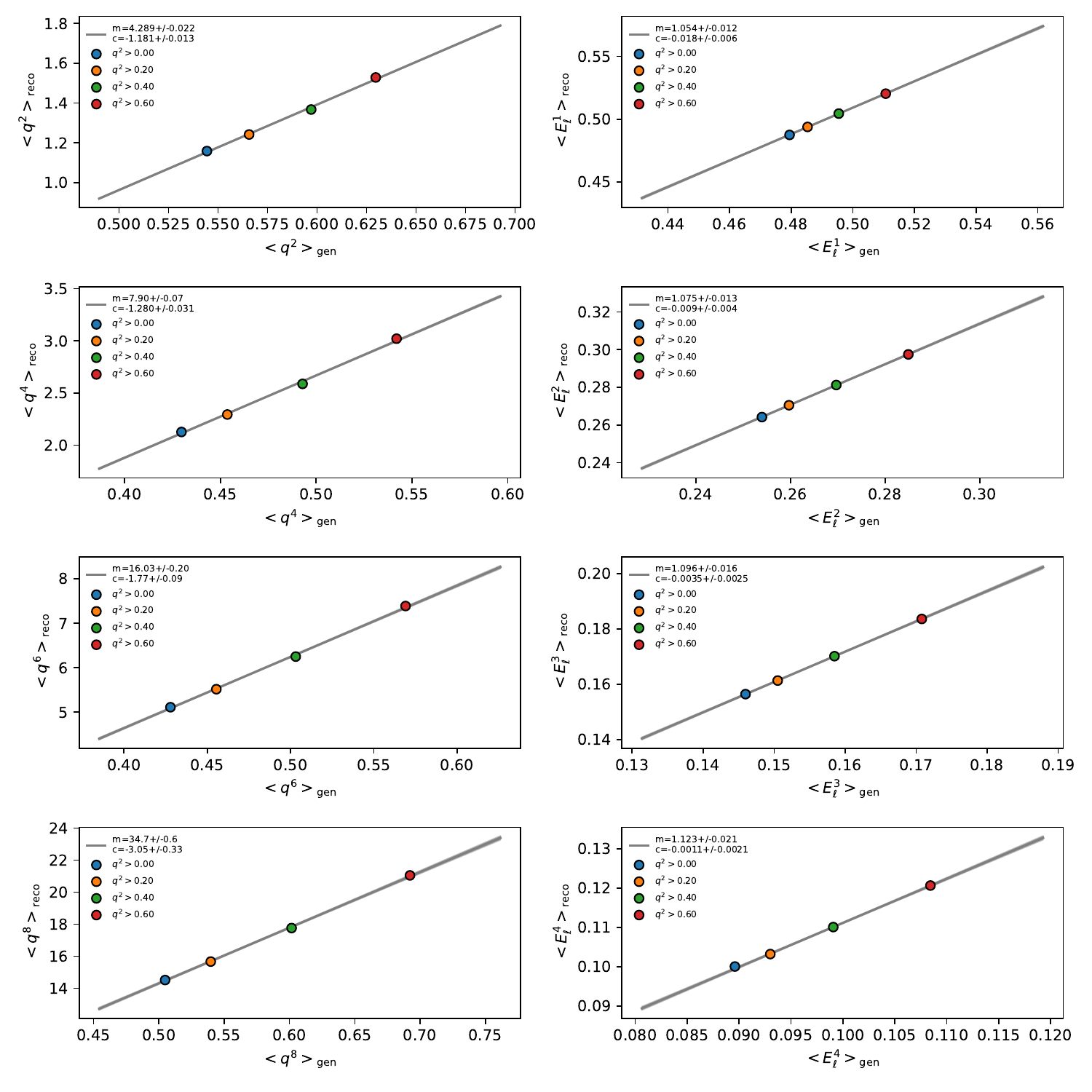}
    \caption{$D_s$ raw moment calibration}
    \label{fig:calibration_Ds}
\end{figure}

\begin{figure}
    \centering
    \includegraphics[width=\linewidth]{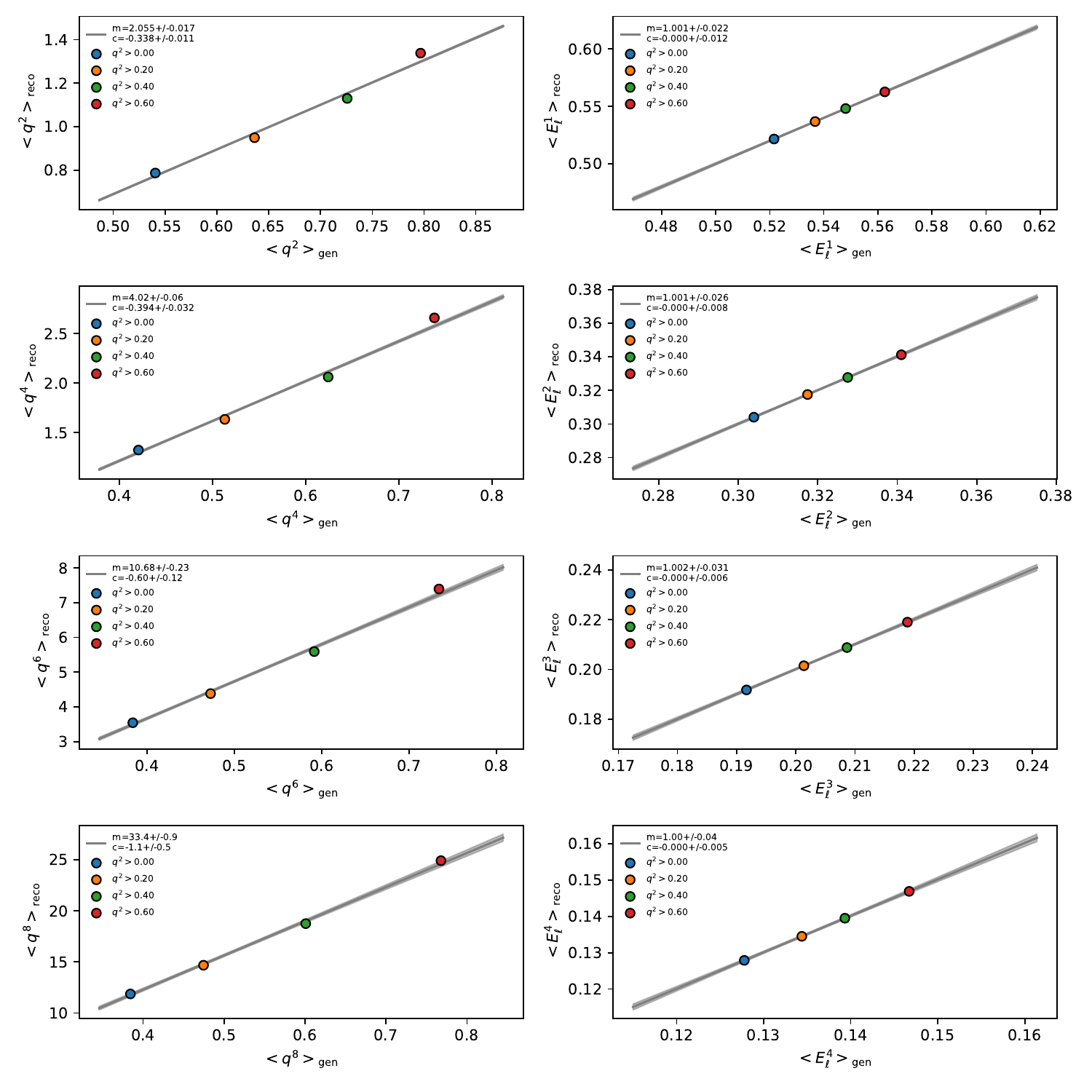}
    \caption{$\Lambda_c$ raw moment calibration}
    \label{fig:calibration_Lc}
\end{figure}

\begin{figure}
    \centering
    \includegraphics[width=\linewidth]{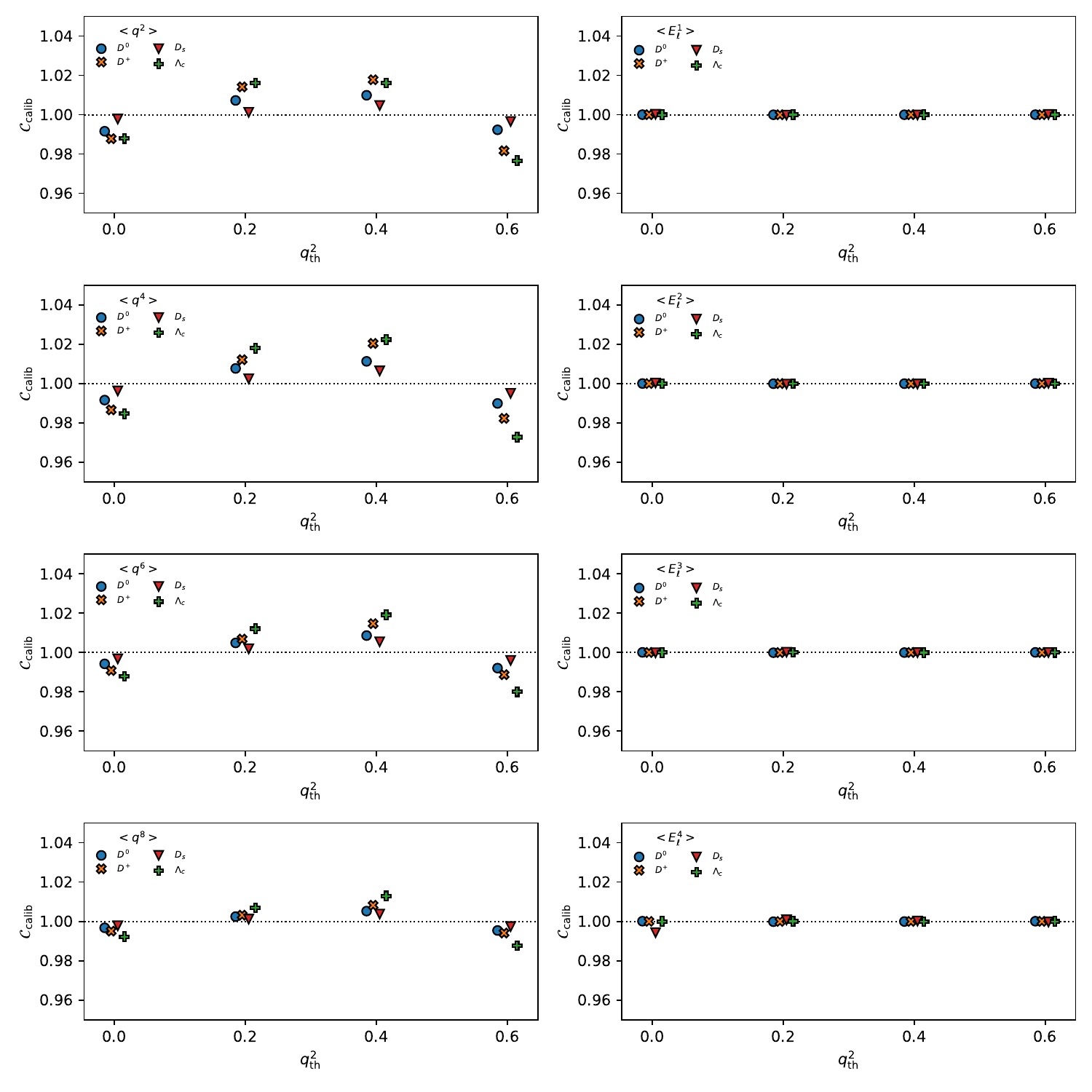}
    \caption{$\mathcal{C}_\mathrm{calib}$ calibration factors.}
    \label{fig:calibration_Calib}
\end{figure}

\begin{figure}
    \centering
    \includegraphics[width=\linewidth]{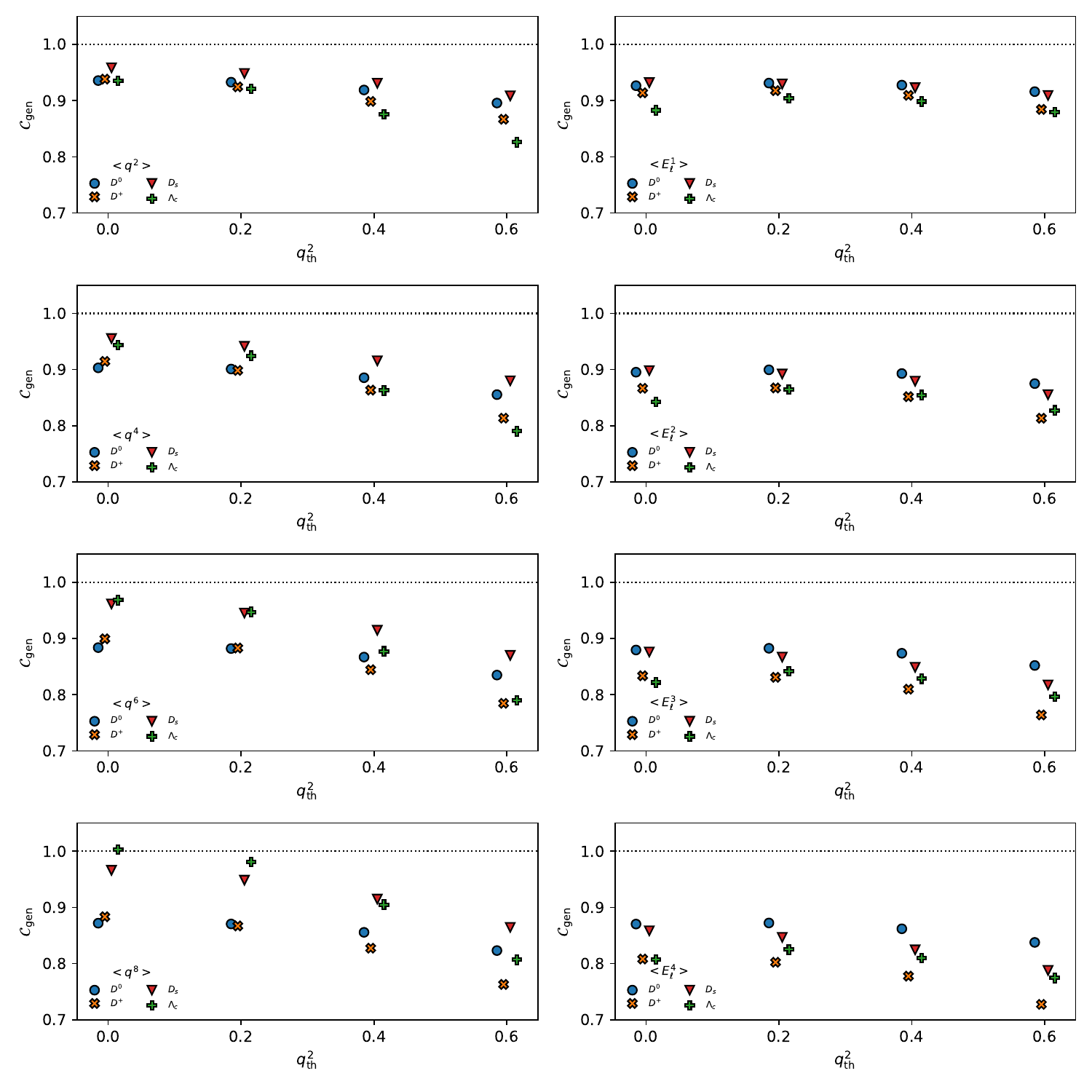}
    \caption{$\mathcal{C}_\mathrm{gen}$ calibration factors.}
    \label{fig:calibration_Cgen}
\end{figure}

The linear calibration functions for the first, second, third, and fourth $q^2$ and $E_\ell$ moments of $D^0$, $D^+$, $D_s^+$ and $\Lambda_c^2$ are depicted in Figures.~\ref{fig:calibration_D0}-\ref{fig:calibration_Lc}. The corresponding $C_{\mathrm{calib}}$ and $C_{\mathrm{gen}}$ corrections for all four states factors are depicted in Figures~\ref{fig:calibration_Calib} and \ref{fig:calibration_Cgen}.

\end{document}

%% file: definitions.tex
\usepackage{xspace} 

\def\Ecm{\ensuremath{E_{\text{cm}}}\xspace}

\def\invfb{\ensuremath{\text{ fb}^{-1}}\xspace}
\def\MeV{\ensuremath{\text{ MeV}}\xspace}
\def\GeV{\ensuremath{\text{ GeV}}\xspace}

\def\KSToPiPi{\ensuremath{K^0_S\to\pi^+\pi^-}\xspace}
\def\KSToNeutrals{\ensuremath{K^0_S\to\pi^0\pi^0}\xspace}
\def\KL{\ensuremath{K^0_L}\xspace}
\def\Dp{\ensuremath{D^+}\xspace}
\def\Dm{\ensuremath{D^-}\xspace}
\def\Dz{\ensuremath{D^0}\xspace}
\def\Dzb{\ensuremath{\overline{D}{}^0}\xspace}
\def\Dsp{\ensuremath{D_{s}^{+}}\xspace}
\def\Dsm{\ensuremath{D_{s}^{-}}\xspace}
\def\Lcp{\ensuremath{\Lambda_{c}^{+}}\xspace}
\def\Lcm{\ensuremath{\Lambda_{c}^{-}}\xspace}
\def\Dz{\ensuremath{D^0}\xspace}
\def\Dzb{\ensuremath{\overline{D}{}^0}\xspace}

\def\DpIncSL{\ensuremath{D^+\to X e^+\nu_e}\xspace}
\def\DzIncSL{\ensuremath{D^0\to X e^+\nu_e}\xspace}
\def\DsIncSL{\ensuremath{D_s^+\to X e^+\nu_e}\xspace}
\def\LcIncSL{\ensuremath{\Lambda_c^+\to X e^+\nu_e}\xspace}

\def\qSq{\ensuremath{q_\text{reco}^2}\xspace}
\def\DqSq{\ensuremath{\Delta q^2}\xspace}
\def\mX{\ensuremath{M_{X}}\xspace}
\def\Elep{\ensuremath{E_{\ell}}\xspace}
\def\Empm{\ensuremath{E_\text{miss}-p_\text{miss}}\xspace}

%% file: tables/moments.tex
\begin{tabular}{lrrrrrrrr}
\toprule
 & $\langle q^{2} \rangle$ & $\langle q^{4} \rangle$ & $\langle q^{6} \rangle$ & $\langle q^{8} \rangle$ & $\langle E_\ell^{1} \rangle$ & $\langle E_\ell^{2} \rangle$ & $\langle E_\ell^{3} \rangle$ & $\langle E_\ell^{4} \rangle$ \\
\midrule
$D^0$ & $1.2$ & $3.0$ & $6.1$ & $11.3$ & $0.3$ & $0.4$ & $0.5$ & $0.6$ \\
$D^+$ & $0.6$ & $0.8$ & $1.2$ & $1.6$ & $0.2$ & $0.3$ & $0.4$ & $0.6$ \\
$D_s$ & $10.5$ & $11.7$ & $9.8$ & $8.1$ & $0.9$ & $1.7$ & $2.9$ & $10.4$ \\
$\Lambda_c$ & $5.6$ & $9.9$ & $15.0$ & $18.7$ & $4.2$ & $5.9$ & $7.2$ & $8.3$ \\
\bottomrule
\end{tabular}

%% file: tables/HQE_parameters.tex
\begin{tabular}{lrrrrr}
\toprule
 & $\mu_\pi^2$ & $\mu_G^2$ & $\rho_D^3$ & $\rho_{LS}^3$ & $\tau_0$ \\
\midrule
$D^0$ & $21.4$ & $29.1$ & $1.4$ & $37.5$ & $66.1$ \\
$D^+$ & $8.5$ & $4.5$ & $0.3$ & $6.7$ & $5.3$ \\
$D_s$ & $94.5$ & $101.1$ & $3.1$ & $92.9$ & $94.4$ \\
$\Lambda_c$ & $50.8$ &  & $2.8$ &  & $86.1$ \\
\bottomrule
\end{tabular}

%% file: tables/D0_systematics.tex

\begin{tabular}{lrrrrr}
\toprule
 & $\mu_\pi^2$ & $\mu_G^2$ & $\rho_D^3$ & $\rho_{LS}^3$ & $\tau_0$ \\
\midrule
Full & 21.4 & 29.1 & 1.4 & 37.5 & 66.1 \\
\midrule
Stat. & 9.2 & 5.9 & 0.2 & 9.6 & 10.9 \\
MC Stat. & 14.7 & 16.4 & 0.3 & 22.6 & 27.1 \\
$\epsilon_{\mathrm{track.}}$ & 16.1 & 26.8 & 1.2 & 32.9 & 62.0 \\
$\sigma_{\mathrm{track.}}$ & 0.2 & 0.2 & 0.0 & 0.3 & 0.3 \\
$\epsilon_{K_S^0}$ & 1.8 & 0.8 & 0.0 & 1.6 & 0.9 \\
$\sigma_{K_S^0}$ & 0.0 & 0.0 & 0.0 & 0.0 & 0.1 \\
$\epsilon_{\gamma}$ & 0.7 & 0.8 & 0.1 & 1.1 & 1.7 \\
$\sigma_{\gamma}$ & 0.3 & 0.1 & 0.0 & 0.1 & 0.9 \\
PID & 0.0 & 0.1 & 0.0 & 0.1 & 0.1 \\
$ \mathcal{B}(D^0 \to K_1^- \ell \nu_\ell )$ & 5.1 & 6.0 & 0.0 & 8.1 & 9.6 \\
$ \mathcal{B}(D^0 \to K^{*-} \ell \nu_\ell )$ & 4.1 & 3.4 & 0.6 & 8.7 & 21.0 \\
$ \mathcal{B}(D^0 \to K^- \ell \nu_\ell )$ & 3.6 & 4.5 & 0.2 & 5.4 & 5.0 \\
$ \mathcal{B}(D^0 \to \rho^- \ell \nu_\ell )$ & 6.0 & 12.6 & 0.7 & 13.4 & 30.1 \\
$ \mathcal{B}(D^0\to (K\pi)_{S-\mathrm{wave}} \ell \nu_\ell) $ & 6.8 & 5.6 & 0.1 & 9.1 & 8.9 \\
\bottomrule
\end{tabular}

%% file: tables/Dp_systematics.tex

\begin{tabular}{lrrrrr}
\toprule
 & $\mu_\pi^2$ & $\mu_G^2$ & $\rho_D^3$ & $\rho_{LS}^3$ & $\tau_0$ \\
\midrule
Full & 8.5 & 4.5 & 0.3 & 6.7 & 5.3 \\
\midrule
Stat. & 5.0 & 0.8 & 0.1 & 3.3 & 1.6 \\
MC Stat. & 3.1 & 1.4 & 0.1 & 2.0 & 2.7 \\
$\epsilon_{\mathrm{track.}}$ & 2.3 & 2.9 & 0.2 & 3.6 & 1.1 \\
$\sigma_{\mathrm{track.}}$ & 0.2 & 0.0 & 0.0 & 0.1 & 0.1 \\
$\epsilon_{K_S^0}$ & 3.2 & 1.7 & 0.1 & 2.1 & 1.9 \\
$\sigma_{K_S^0}$ & 0.1 & 0.1 & 0.0 & 0.1 & 0.0 \\
$\epsilon_{\gamma}$ & 1.5 & 0.9 & 0.0 & 0.9 & 0.9 \\
$\sigma_{\gamma}$ & 0.5 & 0.2 & 0.0 & 0.3 & 1.1 \\
PID & 0.0 & 0.0 & 0.0 & 0.0 & 0.0 \\
$ \mathcal{B}(D^+ \to \eta'\ell \nu_\ell )$ & 0.4 & 0.1 & 0.0 & 0.5 & 1.1 \\
$ \mathcal{B}(D^+ \to \eta \ell \nu_\ell )$ & 1.4 & 0.3 & 0.1 & 1.4 & 2.2 \\
$ \mathcal{B}(D^+ \to K^{*-} \ell \nu_\ell )$ & 2.6 & 0.6 & 0.0 & 1.7 & 0.5 \\
$ \mathcal{B}(D^+ \to K^- \ell \nu_\ell )$ & 2.2 & 0.9 & 0.1 & 2.2 & 0.5 \\
$ \mathcal{B}(D^+ \to \omega \ell \nu_\ell )$ & 0.6 & 0.9 & 0.1 & 0.2 & 0.9 \\
$ \mathcal{B}(D^+ \to \pi^- \ell \nu_\ell )$ & 1.6 & 1.3 & 0.0 & 0.5 & 2.3 \\
$ \mathcal{B}(D^+ \to \rho^- \ell \nu_\ell )$ & 0.9 & 1.1 & 0.1 & 1.3 & 0.7 \\
$ \mathcal{B}(D^+\to (K\pi)_{S-\mathrm{wave}} \ell \nu_\ell) $ & 4.1 & 1.8 & 0.0 & 1.5 & 1.3 \\
\bottomrule
\end{tabular}

%% file: tables/Ds_systematics.tex

\begin{tabular}{lrrrrr}
\toprule
 & $\mu_\pi^2$ & $\mu_G^2$ & $\rho_D^3$ & $\rho_{LS}^3$ & $\tau_0$ \\
\midrule
Full & 94.5 & 101.2 & 3.0 & 92.9 & 94.4 \\
\midrule
Stat. & 34.6 & 9.6 & 0.6 & 28.6 & 18.0 \\
MC Stat. & 25.6 & 25.8 & 1.1 & 21.3 & 15.4 \\
$\epsilon_{\mathrm{track.}}$ & 22.5 & 28.1 & 1.0 & 27.6 & 21.1 \\
$\sigma_{\mathrm{track.}}$ & 8.4 & 10.1 & 0.5 & 5.6 & 4.2 \\
$\epsilon_{K_S^0}$ & 8.0 & 6.9 & 0.2 & 6.7 & 12.1 \\
$\sigma_{K_S^0}$ & 8.6 & 10.1 & 0.5 & 5.7 & 4.2 \\
$\epsilon_{\gamma}$ & 7.9 & 4.3 & 0.2 & 11.2 & 2.4 \\
$\sigma_{\gamma}$ & 10.2 & 7.9 & 0.6 & 8.8 & 7.1 \\
PID & 8.6 & 10.0 & 0.5 & 5.7 & 4.2 \\
$ \mathcal{B}(D_s \to \eta'\ell \nu_\ell )$ & 54.2 & 9.8 & 0.2 & 27.8 & 10.6 \\
$ \mathcal{B}(D_s \to \eta \ell \nu_\ell )$ & 33.3 & 90.3 & 2.3 & 61.0 & 86.5 \\
$ \mathcal{B}(D_s \to f_0 \ell \nu_\ell )$ & 14.4 & 3.3 & 0.2 & 8.6 & 3.1 \\
$ \mathcal{B}(D_s \to K^{*-} \ell \nu_\ell )$ & 30.2 & 21.0 & 1.0 & 30.8 & 5.9 \\
$ \mathcal{B}(D_s \to K^- \ell \nu_\ell )$ & 34.8 & 12.8 & 0.2 & 26.7 & 15.5 \\
$ \mathcal{B}(D_s\ \to \phi \ell \nu_\ell )$ & 33.4 & 19.6 & 0.2 & 36.2 & 23.8 \\
\bottomrule
\end{tabular}

%% file: tables/Lc_systematics.tex

\begin{tabular}{lrrr}
\toprule
 & $\mu_\pi^2$ & $\rho_D^3$ & $\tau_0$ \\
\midrule
Full & 50.8 & 2.8 & 86.1 \\
\midrule
Stat. & 37.0 & 1.4 & 46.1 \\
MC Stat. & 27.6 & 0.9 & 24.4 \\
$\epsilon_{\mathrm{track.}}$ & 7.2 & 0.6 & 17.4 \\
$\sigma_{\mathrm{track.}}$ & 7.4 & 0.3 & 10.0 \\
$\epsilon_{K_S^0}$ & 7.2 & 0.3 & 9.8 \\
$\sigma_{K_S^0}$ & 6.9 & 0.3 & 9.8 \\
$\epsilon_{\gamma}$ & 4.4 & 0.2 & 6.3 \\
$\sigma_{\gamma}$ & 6.0 & 0.1 & 4.8 \\
PID & 7.2 & 0.3 & 9.8 \\
$ \mathcal{B}(\Lambda_c \to \Delta^0 \overline{K}^0 \ell \nu_\ell  )$ & 18.6 & 1.5 & 46.8 \\
$ \mathcal{B}(\Lambda_c \to \Lambda^0 \ell \nu_\ell )$ & 10.3 & 0.7 & 17.2 \\
$ \mathcal{B}(\Lambda_c \to n^0 \ell \nu_\ell )$ & 16.0 & 1.6 & 51.3 \\
$ \mathcal{B}(\Lambda_c \to \Sigma^0 \pi^0 \ell \nu_\ell  )$ & 7.7 & 0.4 & 17.1 \\
$ \mathcal{B}(\Lambda_c \to p^+ K^- \ell \nu_\ell )$ & 15.3 & 1.4 & 40.8 \\
\bottomrule
\end{tabular}